\documentclass[review,11pt]{elsarticle}
\usepackage{lineno,hyperref}
\modulolinenumbers[5]
\journal{Combustion and Flame}
\usepackage{epstopdf}
\usepackage{setspace}
\usepackage{graphicx}
\usepackage{color}
\usepackage{rotating}
\usepackage{soul}
\usepackage{geometry}
\usepackage[version=4]{mhchem}

\biboptions{sort&compress}
\usepackage{caption}
\usepackage[table]{xcolor}
\usepackage{multirow}
\usepackage{arydshln}
\usepackage{amsmath,amssymb}
\usepackage{tikz}
\usetikzlibrary{calc,topaths}

\usepackage{chemfig}
\usepackage{xcolor}

\begin{document}

\begin{frontmatter}

\title{Multi-droplet combustion in the presence of homogeneous and heterogeneous atomizations}
\author{Sepehr Mosadegh and Sina Kheirkhah$^{*}$}
\address{School of Engineering, The University of British Columbia, Kelowna, British Columbia, Canada, V1V 1V7}
\cortext[mycorrespondingauthor]{Corresponding author (sina.kheirkhah@ubc.ca)}

\begin{abstract}
The effects of heterogeneous and homogeneous atomizations on combustion dynamics and flame spread rates of multi-droplets are experimentally investigated. The utilized fuel is biodiesel, which is doped with graphene oxide and blended with ethanol. Stable fuel suspensions are prepared. Either one droplet is positioned at the intersection of two thin fibers or an array of three droplets are positioned along a horizontal line and at the intersections of fibers. For the latter configuration, either the droplet at the center or that on the right-hand-side is ignited. Simultaneous and time-resolved flame chemiluminescence and shadowgraphy techniques are utilized for imaging the flame and droplets, respectively. It is observed that, for a single droplet combustion of biodiesel, relatively low frequency (about 10~Hz) dynamics appear in the second half of the droplet life-time, which is due to heterogeneous atomizations. Blending with ethanol leads to homogeneous atomizations with dynamics featuring frequencies smaller than 50~Hz and at the beginning of the droplet lifetime. It is reported that the homogeneous atomizations lead to at least three times larger root-mean-square oscillations of the flame chemiluminescence compared to that for heterogeneous atomizations. For all tested multi-droplet configurations, independent of the graphene oxide doping concentration and blending with ethanol, the most dominant frequency of flame chemiluminescence oscillations is about 10--15~Hz. It is observed that the heterogeneous atomizations lead to a flame spread rate of about 6--8~$\mathrm{mm^2/s}$, with largest values reported for doping with moderate concentration of graphene oxide. Compared to that for heterogeneous atomizations, the flame spread rate for homogeneous atomizations are significantly larger (10--40~$\mathrm{mm^2/s}$), are not sensitive to graphene oxide doping concentration, and feature a positive relation with the root-mean-square of the flame chemiluminescence oscillations. Overall, the results of the present study underline the important role of atomizations on the flame dynamics and how fast the flame spreads among multi-droplet configurations.

\end{abstract}


\end{frontmatter}

\textbf{Novelty and significance statement}

The dynamics of multi-droplets combustion and the rate of flame spread among the droplets during homogeneous and heterogeneous atomization events are investigated experimentally for the first time. Compared to the heterogeneous atomizations (which occur due to doping with nanomaterials), homogeneous atomizations (which occur due to blending with volatile compounds) significantly influence the flame dynamics and increase the flame spread rate. This study provides insight into the group droplet combustion, which is important for bridging the knowledge gap between single droplet combustion and spray flames.

\textbf{Authors contributions}

SM: Designed and performed the experiments, reduced and analyzed the data, and wrote the manuscript.

SK: Acquired funding, administered the project, supervised the first author, and wrote the manuscript.

\section{Introduction}
\label{Introduction}

Biodiesel doped with nanomaterials and blended with volatile compounds is a popular alternative to traditional diesel due to forming stable suspensions as ``drop-in" fuels, reducing the emission of pollutants from internal combustion engines, and increasing the generated power from these engines~\cite{mahmudul2017production,hoseini2020performance,kwanchareon2007solubility,chao2019microexplosion}. Among the doping nanomaterials studied in the literature, Graphene Oxide (GO) stands out for its thermo-physical properties, notable energy content, absence of toxic metal-oxides during combustion, and the flexibility to incorporate various chemical functional groups with its structure~\cite{ghamari2016experimental,lee2021comprehensive,kegl2021nanomaterials,mosadegh2022graphene,mosadegh2022role}. Either doping with nanomaterials or blending with volatile compounds (e.g. ethanol) can lead to the fuel droplets' atomization~\cite{basu2016combustion,law1982recent}. This phenomenon arises due to the buildup of internal pressure within the droplets, caused by nucleation around nanomaterials within the droplets or the confinement and superheating of volatile species~\cite{law1982recent,basu2016combustion}. Despite several studies investigated the combustion characteristics of doped with nanomaterials or blended with volatile compounds single droplets, multi-droplet configuration is of more relevance to engineering applications, yet our understanding of how such doping or blending (and as a result the corresponding atomizations) influences the combustion characteristics of multi-droplets remains to be developed. In the following, first, the background related the to the atomization phenomenon is reviewed. Then, for single droplets, the background related to the impact of the atomization on combustion dynamics is discussed. Finally, the review of literature concerning the characteristics of flame spread in multi-droplet configurations is presented. \par

\subsection{The droplet atomization phenomenon}
The precursor to the atomization of droplets is the nucleation phenomenon, with two types of nucleation identified in the literature and elaborated below. The first type is referred to as the heterogeneous nucleation and occurs at the liquid-solid interface. The solid interface is either the droplet suspending mechanism, the doping nanomaterial, or both~\cite{ghamari2017combustion,basu2016combustion,mosadegh2022graphene}. For the heterogeneous nucleation, the flame radiative heat is absorbed by the surface of the nanomaterials, causing the surrounding liquid to superheat, which leads to the nucleation. The second type of nucleation (referred to as homogeneous nucleation) occurs at the liquid-liquid interface of multi-component fuels with a volatility mismatch among the constituent components~\cite{meng2021study,meng2022effect}. During the droplet combustion of multi-component fuels, the concentration of the more volatile component at the droplet surface decreases since the more volatile component evaporates faster. This increases the concentration of the less volatile component at the droplet surface. The less volatile component is then heated to its boiling temperature, surpassing that of the more volatile component. As a result, the more volatile component is superheated, nucleation initiates, and bubbles are formed. Whether caused by the heterogeneous or homogeneous nucleation, the generated bubbles may merge, follow the droplet internal-flow pattern, and reach the droplet surface. This leads to the droplet surface rupture and ligaments formation~\cite{basu2016combustion}. Finally, the ligaments might breakup, and as a result, baby droplets and fuel vapor eject out of the droplet~\cite{gan2011combustion,gan2012combustion,miglani2014insight,miglani2015coupled,pandey2019high,miglani2015effect,law2010combustion}. The above described phenomenon is referred to as the atomization \cite{law1982recent}. \par

Heterogeneous and homogeneous nucleations can influence the characteristics of the atomization events. For heterogeneous nucleation caused by the addition of a doping agent, both the concentration and the chemical composition of the doping agents influence the atomization events  \cite{miglani2014insight}. For relatively moderate doping concentrations, the nanomaterials form a shell at the droplet surface that is continuously ruptured during the atomization events~\cite{miglani2015coupled}. However, for excessively large doping concentrations, the shell is relatively strong, which traps the bubbles, and suppresses the atomization events. For GO-doped ethanol droplets, Mosadegh~\textit{et al.}~\cite{mosadegh2022graphene} showed that altering the oxygen content in the molecular structure of GO influenced the amount of infrared light absorbed by this doping agent which affected the corresponding droplets' rate of burning. For homogeneous nucleation and atomization, both the relative volatility and the concentration of the constituent components influence the atomizations~\cite{law1982recent}. Specifically, increasing the difference between the volatility of the constituent components and decreasing the difference between the concentrations of the constituent components lead to larger superheating, increased rate of homogeneous nucleation, and as a result, the occurrence of more pronounced atomization events~\cite{law1982recent}. In the present study and for brevity, the atomization caused by heterogeneous (homogeneous) nucleation is referred to as heterogeneous (homogeneous) atomization. \par

\subsection{Influence of atomization on single droplet combustion dynamics}
The combustion dynamics of a single droplet during homogeneous atomizations was investigated by Pandey~\textit{et al.}~\cite{pandey2019high}. They~\cite{pandey2019high} showed that, for ethanol mixed with water droplets, the droplet shape and the flame chemiluminescence featured oscillations at about 32 and 43~Hz, respectively. For heterogeneous atomizations, the concentration of the doping agent significantly influenced the dynamics, see for example~\cite{miglani2014insight,miglani2015effect,pandey2019high,mosadegh2022role}. For relatively dilute doping concentrations of the nanomaterials, Miglani~\textit{et al.}~\cite{miglani2015effect} showed that the atomizations led to small-scale droplet deformation. However, for relatively large doping concentrations, they~\cite{miglani2015effect} showed that relatively intense ejections occurred which led to large amplitude droplet surface deformations. Independent of the doping concentration, Miglani~\textit{et al.}~\cite{miglani2014insight,miglani2015effect} concluded that the bubbles to droplet volume ratio significantly influenced the breakup, the atomizations, and as a result, the droplet surface deformation. 

Motivated by understanding the influence of heterogeneous atomizations on the relation between the droplet shape deformation and flame chemiluminescence oscillations, Mosadegh and Kheirkhah~\cite{mosadegh2022role} investigated the dynamics of diesel droplets doped with GO. It was shown~\cite{mosadegh2022role} that the doped droplet size and flame chemiluminescence featured intermittent oscillations. It was also reported that increasing the doping concentration increased both the mean number and the intensity of the atomization events, leading to an increase in the root-mean-square (rms) of the droplet diameter squared oscillations~\cite{mosadegh2022role}. For all doping concentrations tested in Mosadegh and Kheirkhah~\cite{mosadegh2022role}, the dominant frequencies of the droplet size and flame chemiluminescence oscillations were relatively small (less than or about 15~Hz), and these frequencies locked onto the frequency of the atomization events. \par

\subsection{Influence of atomization on flame spread among several droplets}
The spread of flame among several droplets positioned along a straight line has been investigated in several past studies, see for example~\cite{mikami2005microgravity,mikami2006microgravity,nomura2011microgravity,mikami2018flame,wang2021flame,yamada2015comparison,wang2023micro}. Review of the literature suggests that, among many parameters (such as the background gas composition and temperature \cite{mikami2006microgravity,nomura2011microgravity}), the inter-droplet spacing normalized by the droplet initial diameter ($S/D_0$) substantially influenced the modes of propagation and the flame spread rate ($SD_0/\tau_\mathrm{s}$), with $\tau_\mathrm{s}$ (referred to as the spread time) being the time it takes for the side droplet to be enveloped by the flame~\cite{mikami2005microgravity,mikami2006microgravity}. For a relatively small inter-droplet spacing normalized by the initial droplet diameter ($S/D_0 \lesssim 2$), Mikami \textit{et al.}~\cite{mikami2006microgravity} showed that, as the visible flame size of the main droplet increased, the side droplet was enveloped by the flame. This was followed by the evaporation of the side droplet and its ignition. This mode of flame spread was referred to as Mode~1 in Mikami~\textit{et al.}~\cite{mikami2006microgravity}. For relatively larger inter-droplet spacing ($2 \lesssim S/D_0 \lesssim 6$), the flame of the main droplet reached the flammable mixture layer that is formed around the side droplet and the flame propagated inside the layer and around the side droplet. This was referred to as Mode~2 of flame spread~\cite{mikami2006microgravity}. For significantly larger inter-droplet spacing ($6 \lesssim S/D_0 \lesssim 14-17$) and depending on the background gas temperature, the side droplet may auto-ignite due to heat transfer from the main droplet flame, which cannot reach the flammable mixture layer around the side droplet. This mode of flame spread was referred to as Mode~3 in~\cite{mikami2006microgravity}. For $S/D_0 \gtrsim 17$, the flame did not spread among the droplets~\cite{mikami2006microgravity}.

Although the above investigations are of significant importance as they provided insight into the modes of flame spread in a multi-droplet configuration, the studies were performed for fuels or test conditions that did not yield droplet atomization. Understanding the effect of atomization on the flame spread rate in a multi-droplet configuration is important, as the majority of the fuels used in engineering applications are blends of components with different volatilities, which promoted atomizations. To the authors best knowledge, only the studies of Yamada~\textit{et al.}~\cite{yamada2015comparison} and Wang~\textit{et al.}~\cite{wang2023micro} aimed to relate the rate of flame spread in a multi-droplet configuration to the atomizations. Specifically, Yamada~\textit{et al.}~\cite{yamada2015comparison} investigated the flame spread rate among the droplets of n-dodecane and water emulsions. They~\cite{yamada2015comparison} showed that the flame spread rate was influenced by the water concentration. Wang~\textit{et al.}~\cite{wang2023micro} measured the flame spread rate among the droplets of Jatropha oil (JO) blended with 5-dimethylfuran (DMF). They~\cite{wang2023micro} showed that, for a given inter-droplet spacing, the addition of DMF (that is the more volatile component) to Jatropha oil increased the flame spread rate. \par

Given the above review of literature, two gaps of knowledge are identified. First, despite that the effects of heterogeneous and homogeneous atomizations on single droplet combustion dynamics have been studied in the past (see for example \cite{mosadegh2022role,pandey2019high}), the effects of such atomizations on the dynamics of multi-droplets are yet to be investigated. Second, even though past studies~\cite{yamada2015comparison,wang2023micro} reported the presence of atomization events during the spread of the flame in a multi-droplet configuration and observed that the atomizations can influence the flame spread, the effects of heterogeneous and homogeneous atomizations on the rate of flame spread were not investigated. Thus, the first objective of this work is to study the effects of heterogeneous and homogeneous atomizations on the dynamics of multi-droplet combustion. The second objective of this study is to investigate the effects of heterogeneous and homogeneous atomizations on the flame spread rate in a multi-droplet configuration.

\section{Experimental methodology}
\label{Methodology}
\subsection{Tested fuels}
\label{subsec:testedfuels}
Graphene oxide nanomaterials functionalized with octadecyl amine groups were synthesized by our industrial partner and were utilized as the doping agent in the present study. The detailed characterization of the utilized graphene oxide is provided in the supplementary materials. The GO nanomaterials were added to either biodiesel or 60\% by mass of biodiesel blended with 40\% by mass of ethanol. In total, three GO doping concentrations of 0, 0.01, and 0.1\% (by mass) were examined. A summary of the tested fuels are presented in Table~\ref{tab:Conditions_B}. The first column presents the acronyms for the fuels. In this column, B refers to the tested fuel being biodiesel, and BE refers to the tested fuel being ethanol blended with biodiesel. For tested fuels denoted by B and BE, nanomaterials were not added to the liquid fuel. B$X$ refers to GO added to biodiesel with a doping concentration of $X$\% (by mass), and BE$X$ refers to $X$\% (by mass) of GO added to ethanol blended with biodiesel, with $X = 0.01$ and 0.1. The second and third columns show the mass percentage of biodiesel and ethanol in the blends, respectively. Finally, the fourth and fifth columns are the type of the added nanomaterial and the tested doping concentration (by mass), respectively. \par

\begin{table} [h!] 
	\caption {Test fuels. All percentages are calculated based on the mass ratios. B and BE refer to biodiesel and biodiesel blended with ethanol, respectively. \label{tab:Conditions_B}}
\begin{center}

\scalebox{1.05}{
 
	\begin{tabular}{c c c c c}
			\hline
			\hline
			Tested fuels & $\mathrm{[Biodiesel]}$\%& $\mathrm{[Ethanol]}$\%& Nanomaterials & $\mathrm{[GO]}$\%\\
			\hline
			B      & 100 & 0  & ---    & 0    \\
			BE     & 60  & 40 & ---  & 0         \\
			B0.01  & 100 & 0  & GO  & 0.01   \\
			BE0.01 & 60  & 40 & GO  & 0.01   \\
			B0.1   & 100 & 0  & GO  & 0.1     \\
            BE0.1  & 60  & 40 & GO  & 0.1    \\
			\hline
			\hline	
		\end{tabular}}
\end{center}
\end{table}

\begin{figure}[!h]
	\centering
	\includegraphics[width=0.5\textwidth]{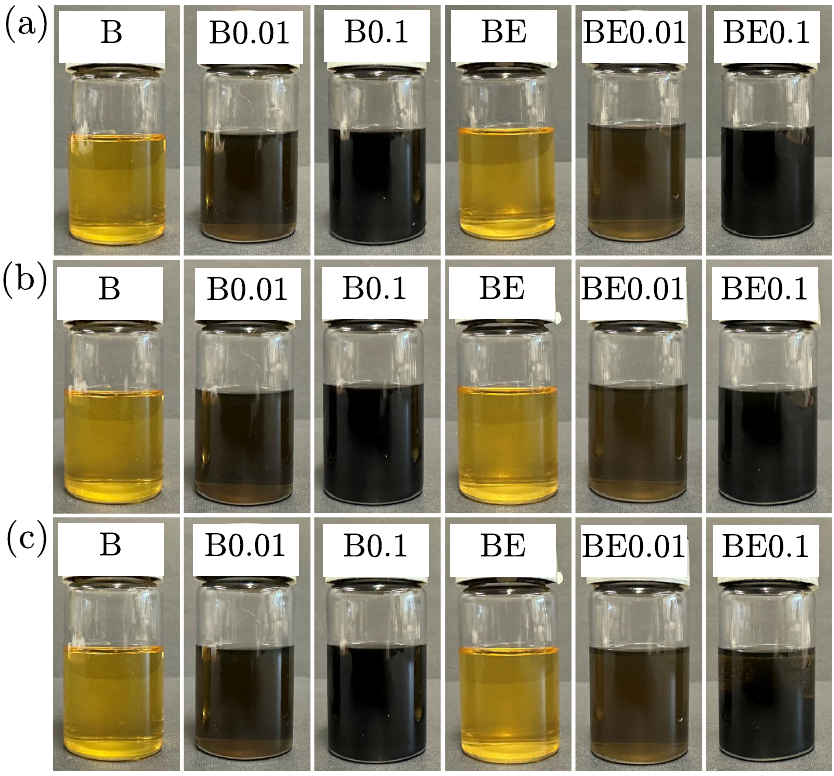} 
	\caption{The pictures of the fuel vials. The pictures in (a), (b), and (c) were acquired immediately, 40~min, and 60~min after sonication. The first to sixth columns pertain to B, B0.01, B0.1, BE, BE0.01, and BE0.1, respectively.}
	\label{fig:Biofuel pictures}
\end{figure}

The stability of the fuel suspensions tested in the present study was investigated. The suspensions of GO and the base fuel were sonicated for 60~min prior to each combustion experiment. Figures~\ref{fig:Biofuel pictures}(a), (b), and (c) present the pictures of the vials containing the fuels immediately, 40~min, and 60~min after the sonication, respectively. The first to sixth columns pertain to B, B0.01, B0.1, BE, BE0.01, and BE0.1, respectively. As can be seen, except BE0.1, all doped fuels formed stable suspensions for at least one hour after they were prepared. BE0.1 was stable for at least 40~min. Each experiment was completed during about two mins, and as a result, all tested fuels were stable for durations significantly longer than the combustion experiments. \par

\subsection{Experimental setup and diagnostics}
\label{Setup}

\begin{figure}[!h]
    \centering
    \includegraphics[width=1\textwidth]{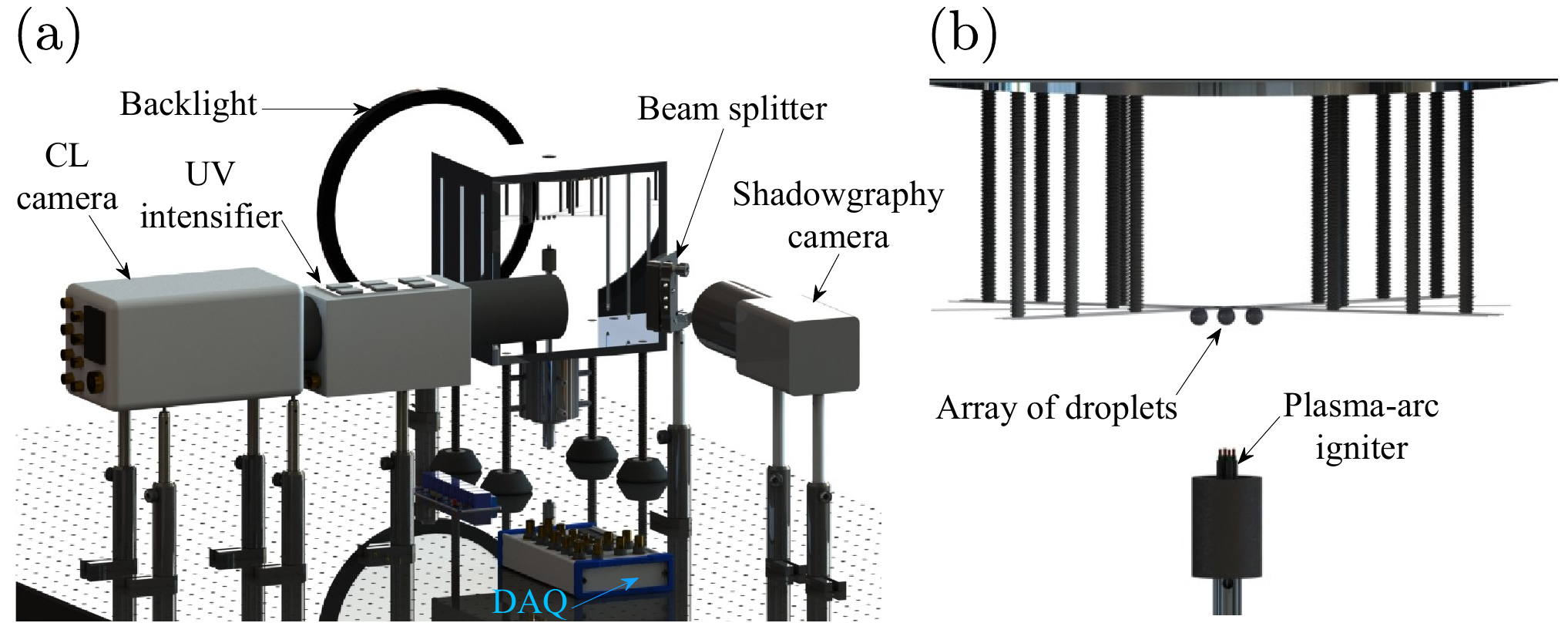}
    \caption{(a) The experimental setup and diagnostics arrangement. (b) is the enlarged view of the setup near the suspended droplets. CL refers to chemiluminescence.}
    \label{fig:setup_B}
\end{figure}

In order to study the combustion dynamics of multi-droplets and estimate the rate of flame spread among them, the experimental setup used in~\cite{mosadegh2022role} was modified to accommodate suspending more than one droplet. The experimental setup used in the present study is shown in Fig.~\ref{fig:setup_B}(a). For each test condition, three 1.9$\pm$0.02~mm diameter droplets were deposited at the intersection of three 140~$\mathrm{\mu}$m SiC fibers. Aiming to maximize the optical accessibility, the angle between the fibers was set to be relatively small, which was $15^\mathrm{o}$ between the fibers. The enlarged view of the droplets suspended on the fibers is shown in Fig.~\ref{fig:setup_B}(b). The inter-droplet spacing was $S = 4$~mm once the droplets were deposited on the fibers. Immediately after the deposition of the droplets, a 20~mW plasma-arc igniter (mounted on a linear actuator) was positioned about 1~mm below a droplet and operated for 400~ms. After the plasma arc operation was complete, the actuator retracted the igniter to minimize its interaction with the flame. Then, the flame enveloped the ignited droplet. This droplet is referred to as the main droplet for the multi-droplet configuration experiments. After the ignition of the main droplet, its flame ignites the side droplet(s). Once the ignition is complete, the main and side droplets may move along the horizontal axis and the inter-droplet spacing may change, which is due to the effect of the recoiling force from the atomizations of the droplets. \par

Compared to our previous studies (see~\cite{mosadegh2022graphene,mosadegh2022role}), in which separate flame chemiluminescence and shadowgraphy techniques were employed, here, simultaneous flame chemiluminescence and shadowgraphy techniques were utilized, with the corresponding arrangement of the diagnostics shown in Fig.~\ref{fig:setup_B}(a). The hardware for the flame chemiluminescence measurements included a Nova S12 high-speed camera, an Invisible Vision UVi 1850B intensifier, a Nikkor lens (focal length of 50~mm) which connected the camera to the intensifier, a UV Nikon Rayfact (focal length of 105 mm and aperture number of 4.5) lens, and a 310$\pm$10~nm band-pass filter. The collected flame chemiluminescence signal features contributions from $\mathrm{OH^*}$, $\mathrm{CO_2^*}$, and $\mathrm{HCO^*}$. Though this signal is not utilized to quantify the heat release rate, it can be used to detect approximate location of the flame and how it spreads between droplets. The collected chemiluminescence intensity was normalized by the white field, and the resultant images were filtered using a median-based filter to remove ``salt-pepper" noise, similar to our previous investigations~\cite{mosadegh2022graphene,mosadegh2022role}. The acquisition frequency for the flame chemiluminescence images was 4000~Hz, and the pixel resolution was 43~$\mu$m. \par

The hardware for the shadowgraphy technique was identical to that used in~\cite{krebbers2023relations}. Specifically, the Andor Zyla 5.5 sCMOS camera was equipped with a Macro Sigma lens (focal length of 105~mm and aperture number of 2.8) to collect the images. Similar to~\cite{mosadegh2022role,bennewitz2019systematic,bennewitz2020combustion}, a Semrock FF01–433/530 200 dual bandpass filter was installed on the shadowgraphy camera lens to improve the visibility of the droplets in the presence of sooty flames. Similar to~\cite{sim2018effects}, a beam splitter with $45^\mathrm{o}$ orientation was used to reflect the light with wavelength smaller than 350~nm for flame chemiluminescence imaging and to transmit light with wavelength larger than 350~nm for shadowgraphy imaging. The acquisition frequency of the shadowgraphy was limited by the utilized camera and was 80~Hz. Though relatively small, this acquisition frequency was sufficient for measuring the large scale motion of the droplets. The shadowgraphy pixel resolution was 13~$\mu$m. Following the procedure elaborated in~\cite{mosadegh2022graphene}, a USAF target was used to obtain the resolving limits of the flame chemiluminescence and shadowgraphy imaging, which were 78 and 14~$\mathrm{\mu}$m/pixel, respectively.

For each tested fuel presented in subsection~\ref{subsec:testedfuels}, the experiments were performed for both single and multi-droplet configurations. For the latter, since three droplets are positioned along a horizontal line, two ignition scenarios were separately examined: the droplet at the rightmost position and the droplet at the center were ignited. In total, 6 fuels listed in Table~\ref{tab:Conditions_B}, 3 droplet configurations (one single droplet and two multi-droplet configurations), and 12 repeats of each condition, which equals $6\times3\times12=216$ experiments were performed.

\section{Results}
\label{results}
The influences of atomization on the combustion dynamics and flame spread rate are discussed in the first and second subsections, respectively. Of importance to the discussions presented below are the procedures for reducing the acquired shadowgraphy and flame chemiluminescence data, with the details of the data reduction presented in the supplementary materials.\par

\subsection{Effect of atomization on combustion dynamics}
\label{subsec:dynamics}
For comparison purposes, first, the dynamics of single droplet combustion is discussed in subsection~\ref{subsec:singledynamics}. Then, the dynamics of the multi-droplet combustion is presented in subsection~\ref{subsec:multidynamics}. \par

\subsubsection{Single droplet combustion dynamics}
\label{subsec:singledynamics}
\begin{figure}[h]
\centering
\includegraphics[width=1\textwidth]{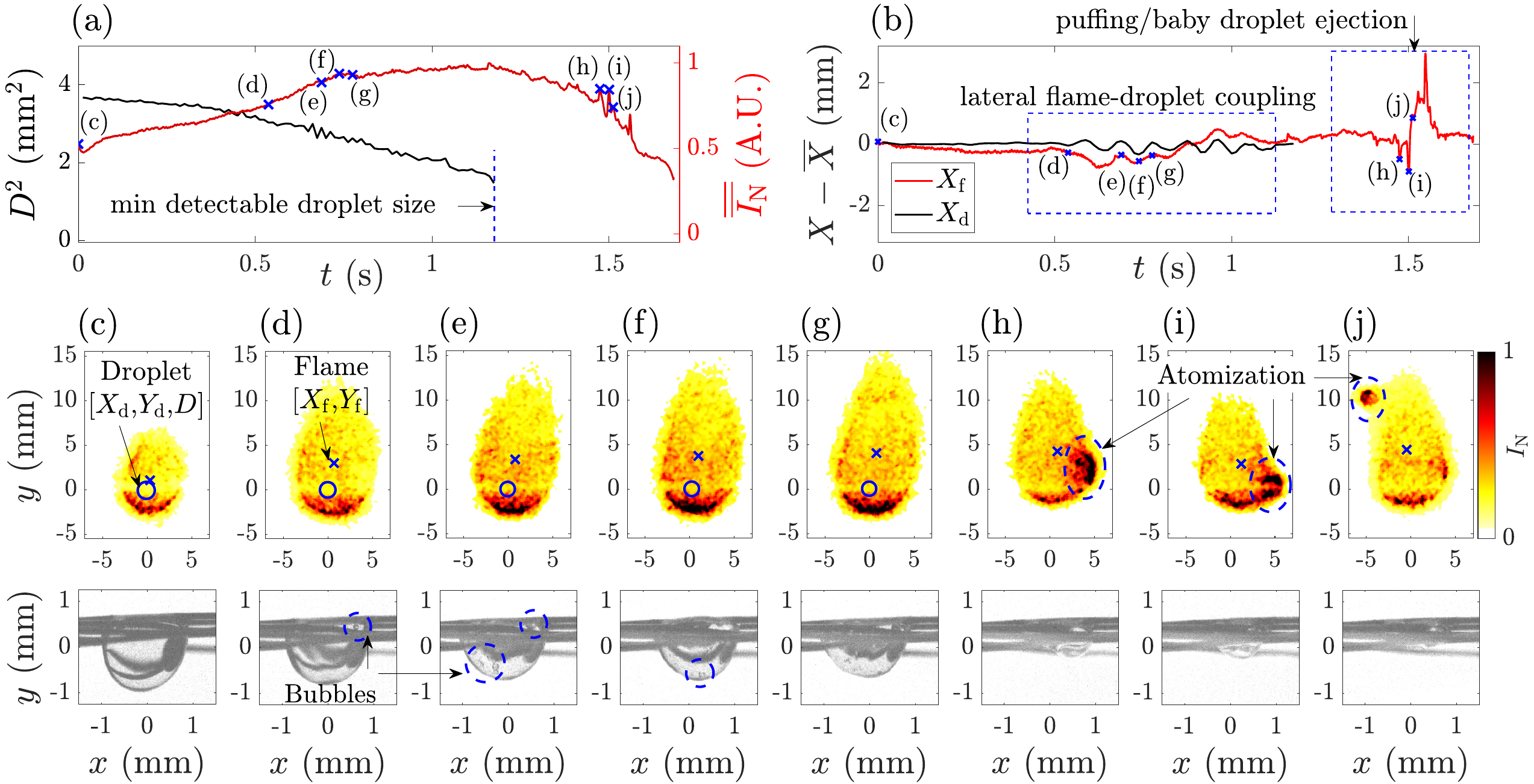}
\caption{(a) Variations of the droplet diameter squared (black) and normalized flame chemiluminescence (red) versus time. (b) Variations of the horizontal position of the droplet (black) and flame chemiluminescence (red) centroids subtracted by their corresponding mean values versus time. (c--j) are the normalized flame chemiluminescence (top panels) and the droplet shadowgraphy images (bottom panels) corresponding to the instants marked in (a) and (b). Specifically, the results in (c--j) correspond to $t = 0$, 0.54, 0.69, 0.74, 0.78, 1.47, 1.50, and 1.51~s, respectively. The results are for the representative test condition of B0.01.}
\label{fig:B100Regimes}
\end{figure}

Key to understanding the dynamics of single droplet combustion is the temporal variation of the droplet diameter squared, $D^2$, versus time. Our analyses show that, compared to graphene oxide, such dynamics are strongly dependent on the presence of ethanol in the fuel. Thus, first, the dynamics of biodiesel doped with GO and then that for ethanol blended with biodiesel and doped with GO are discussed. For a single biodiesel droplet doped with 0.01\% of graphene oxide, the variation of $D^2$ versus time is presented in Fig.~\ref{fig:B100Regimes}(a) using the black curve. Analysis of the shadowgraphy images show that the droplet size becomes comparable to the size of the suspension fibers towards the end of the droplet lifetime, and as a result, the droplets cannot be discerned from the fibers. The minimum detectable droplet size from the shadowgraphy images is shown in Fig.~\ref{fig:B100Regimes}(a). For comparisons, the local flame chemiluminescence, $I_\mathrm{f}(x,y,t)$, was spatially integrated and normalized by its maximum, with this normalized parameter referred to as $\overline{\overline{I_\mathrm{N}}}(t)$. The variation of $\overline{\overline{I_\mathrm{N}}}(t)$ versus time is presented in Fig.~\ref{fig:B100Regimes}(a), using the red curve. As suggested in~\cite{mosadegh2022role}, in addition to the droplet diameter and flame chemiluminescence, the horizontal locations of the droplet and the flame chemiluminescence centroids facilitate understanding the droplet combustion dynamics. The shadowgraphy images were binarized, and the variation of the horizontal location of the droplet centroid subtracted by its mean ($X_\mathrm{d}-\overline{X_\mathrm{d}}$) versus time is presented in Fig.~\ref{fig:B100Regimes}(b) using the black curve. The horizontal location of the flame chemiluminescence centroid, $X_\mathrm{f}(t)$ was calculated using

\begin{equation}
	\label{Eq:xf}
	X_\mathrm{f}(t) = \frac{\int_{x_{\min}}^{x_{\max}}\int_{y_{\min}}^{y_{\max}} xI_\mathrm{f}(x,y,t)\mathrm{d}x\mathrm{d}y}{\int_{x_{\min}}^{x_{\max}} \int_{y_{\min}}^{y_{\max}} \mathrm{d}x\mathrm{d}y},
\end{equation}
with $(x_{\min},x_{\max})$ and $(y_{\min}$,$y_{\max})$ being the horizontal and vertical extents of the acquired flame chemiluminescence data. The variation of $X_\mathrm{f}(t)$ subtracted by its mean value is also presented in Fig.~\ref{fig:B100Regimes}(b) using the red curve. It is important to note that, for $t<0$, once the ignition mechanism is turned off, a small flame kernel is formed at the top of the droplet; and, this kernel propagates around the droplet. $t=0$ is the moment at which the flame kernel has just enveloped the droplet, and, this moment is defined as when the droplet is ignited. For $t = 0$, 0.54, 0.69, 0.74, 0.78, 1.47, 1.50, and 1.51~s, the flame chemiluminescence images normalized by their respective maximum are presented in the top panels of Figs.~\ref{fig:B100Regimes}(c--j) and the corresponding shadowgraphy images are shown in the bottom panels. To aid the discussions, circles with diameter of $D$ (details of calculation are provided in the supplementary materials) are superimposed on Fig.~\ref{fig:B100Regimes}(c--g). The blue cross symbols present the centroid of the flame chemiluminescence in the figures. \par

Comparison of the results in Fig.~\ref{fig:B100Regimes}(c) and those in Figs.~\ref{fig:B100Regimes}(d--g) show that, after the droplet is ignited, the area corresponding to the detected flame chemiluminescence increases in time, which is due to the vaporization of the droplet~\cite{pandey2019high}. As the droplet combustion proceeds, bubbles are formed at the suspending fiber or at the GO aggregates. Such bubble formation is a characteristic of heterogeneous nucleation~\cite{basu2016combustion}, with the images of the bubbles highlighted in Figs.~\ref{fig:B100Regimes}(d--f) using the dashed blue circles. When these bubbles reach the droplet's surface, it ruptures, leading to atomizations during $0.5~\mathrm{s} \lesssim t \lesssim 0.8~\mathrm{s}$. The results in Fig.~\ref{fig:B100Regimes}(a) show that, for such atomizations, the droplet diameter and the spatially integrated flame chemiluminescence do not feature significant fluctuations. Compared to these, the results presented in Fig.~\ref{fig:B100Regimes}(b) show that, during the above atomization, the centroids of both the droplet and flame move laterally and almost synchronously (see the red and black curves corresponding to $0.5~\mathrm{s}\lesssim t \lesssim 1.0~\mathrm{s}$). The lateral movement of the droplet is due to the recoiling force~\cite{pandey2019high} induced as a result of heterogeneous atomizations. This mechanistic relation was hypothesized in \cite{mosadegh2022role} using separate chemiluminescence and shadowgraphy imaging, and the simultaneous flame and droplet measurements of this study confirm the hypothesis in~\cite{mosadegh2022role}. Towards the end of the droplet lifetime, a few intense atomizations occur as shown in Figs.~\ref{fig:B100Regimes}(a and b), with the relevant flame chemiluminescence images presented in Figs.~\ref{fig:B100Regimes}(h--j) and the corresponding changes in flame topology highlighted by the dashed blue ellipses. \par

\begin{figure}[h]
	\centering
	\includegraphics[width=1\textwidth]{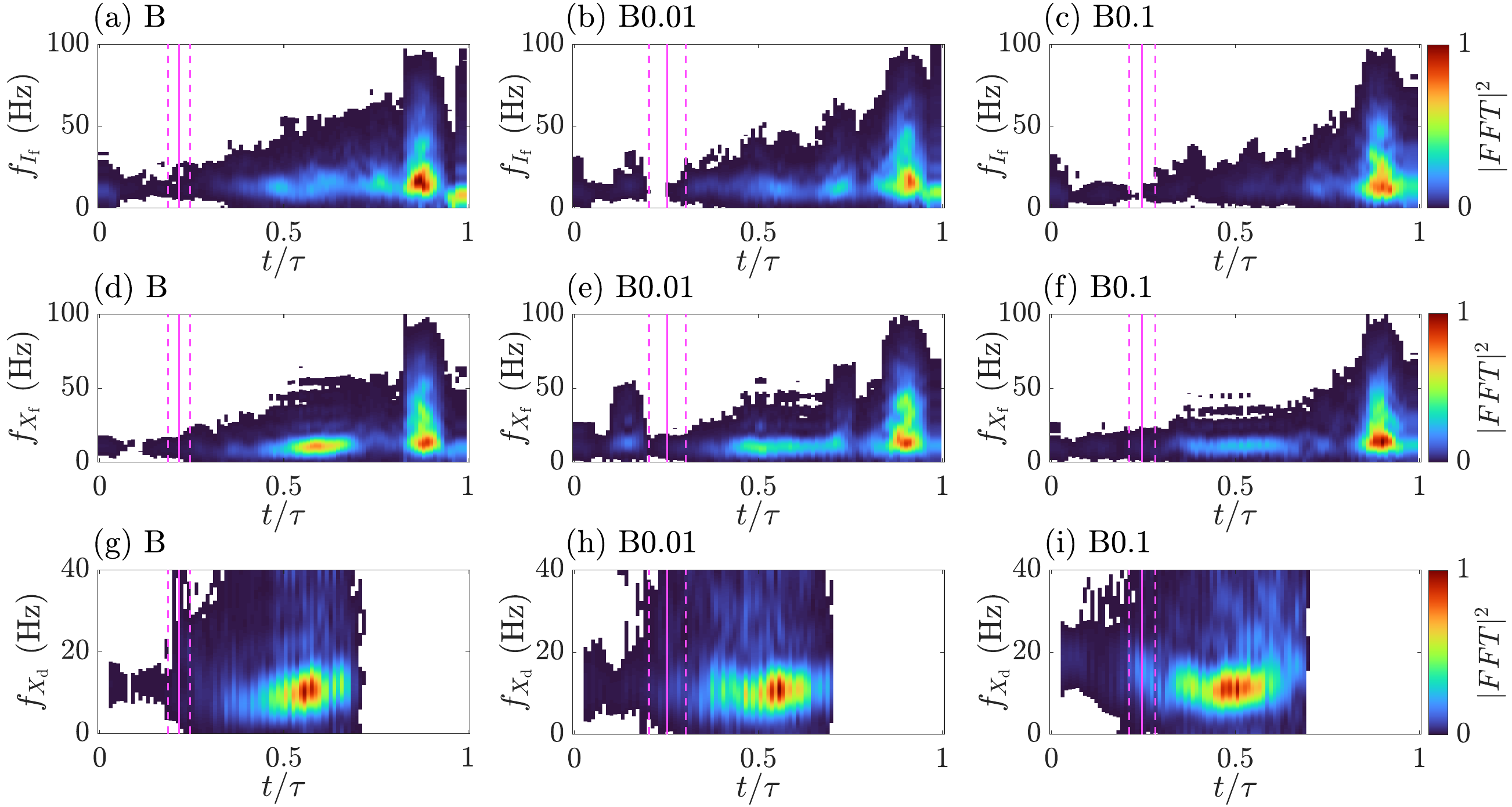}
	\caption{The spectrograms of the doped biodiesel single droplet combustion. (a--c) are for the spatially integrated flame chemiluminescence, (d--f) are for the horizontal location of the flame chemiluminescence centroid, and (g--i) are for the horizontal location of the droplet center. The results in the first, second, and third columns are for biodiesel, biodiesel doped with 0.01\% of GO, and biodiesel doped with 0.1\% of GO, respectively. The solid magenta line is the normalized time at which the heterogeneous atomizations start.}
	\label{fig:Bsingle}
\end{figure}

Aiming to understand the dynamics of the above described oscillations during the atomizations, the spectrograms of $I_\mathrm{f}$, $X_\mathrm{f}$, and $X_\mathrm{d}$ are calculated and presented in Figs.~\ref{fig:Bsingle}(a--c), (d--f), and (g--i), respectively. The results in the first, second, and third columns of the figure correspond to test conditions for which the fuel was biodiesel, biodiesel doped with 0.01\% by mass of GO, and biodiesel doped with 0.1\% by mass of GO, respectively. To obtain the spectrograms, the Fast Fourier Transform (FFT) raised to the power of two of the corresponding data was calculated over a temporal window of 0.2~s. The horizontal axis of the spectrograms present time normalized by the droplet lifetime ($t/\tau$). The spectrograms were calculated for all repeats of each test condition, were averaged over the repeats, and the results were normalized by the maximum. Overlaid on the spectrograms are the straight solid lines in magenta color, which is the mean of the normalized time at which the atomizations start. This time is extracted from the shadowgraphy images. The standard deviation of the above time was calculated based on the 12 repeats of the experiments. The mean normalized times at which the atomizations start plus and minus half of the standard deviation are also presented using the magenta color dashed lines.

It is important to note that, since the shadowgraphy images do not allow for visualizing the droplets after about 70\% of the droplet lifetime, and also, the frequency of the shadowgraphy imaging is relatively small (80~Hz), comparisons between the spectral content of $X_\mathrm{d}$ with those of $I_\mathrm{f}$ and $X_\mathrm{f}$ can only be made for $t/\tau \lesssim 0.7$ and $f \lesssim 80/2 = 40$~Hz. Given the above, several conclusions can be drawn from the results in Fig.~\ref{fig:Bsingle}. First, for single biodiesel droplets (with and without the doping agent), the spectral content of $I_\mathrm{f}$ and $X_\mathrm{f}$ are similar, suggesting that the dynamics of the flame chemiluminescence and that of the droplet movements along the horizontal axis is coupled, which confirms the results presented in Fig.~\ref{fig:B100Regimes}(b). For $t/\tau \lesssim 0.7$ and $f \lesssim 80/2 = 40$~Hz, the spectral content of $X_\mathrm{d}$ is similar to those of $I_\mathrm{f}$ and $X_\mathrm{f}$. Specifically, after about 20\% of the droplet lifetime (at which point the heterogeneous atomizations start), both the flame and the droplet dynamics feature low-frequency (less than 10~Hz) oscillations for all tested fuels. Towards the end of the droplet lifetime ($t/\tau \gtrsim 0.8$), $I_\mathrm{f}$ and $X_\mathrm{f}$ feature large amplitude oscillations for $10 \lesssim f \lesssim 100$~Hz.

\begin{figure}[h]
\centering
\includegraphics[width=1\textwidth]{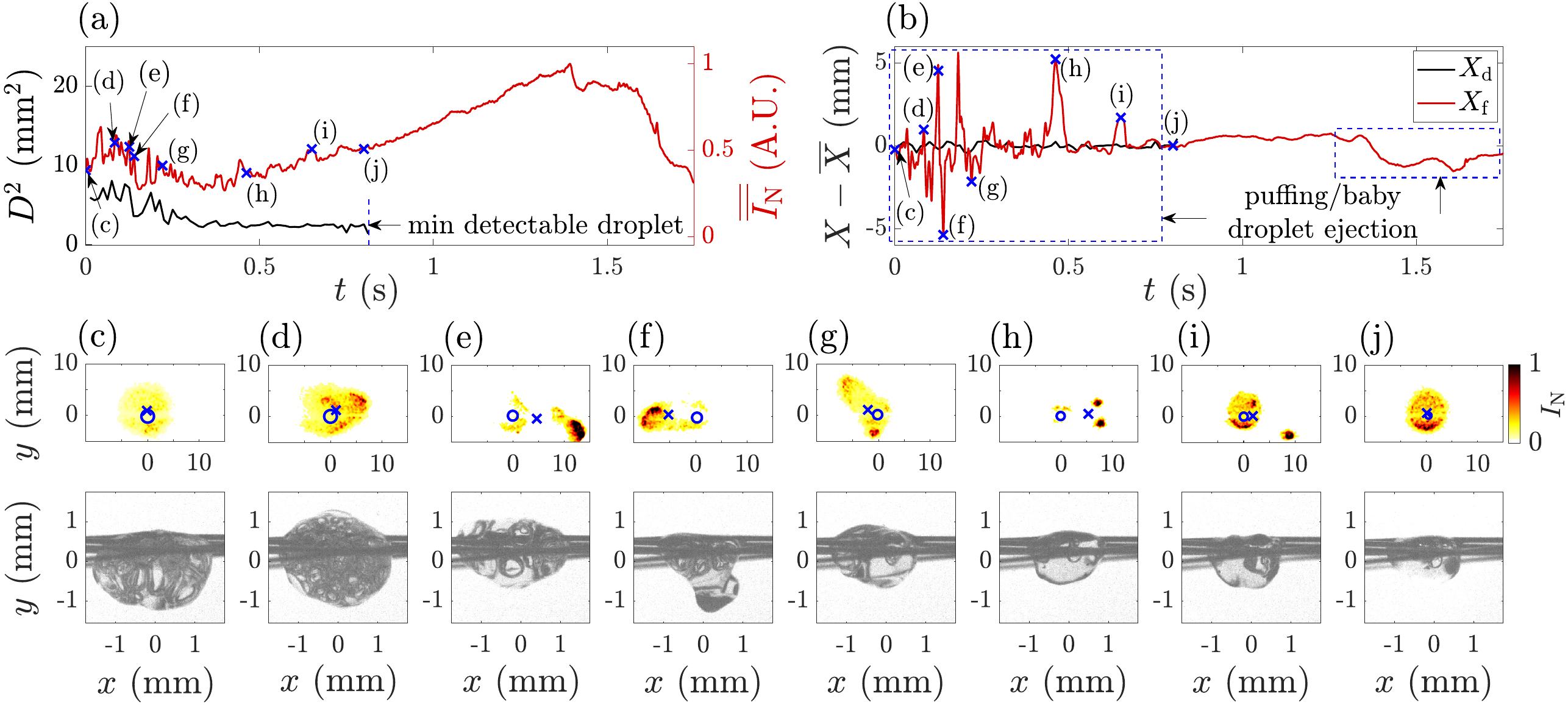}
\caption{(a) The droplet diameter squared (black) as well as the spatially averaged and normalized flame chemiluminescence (red) versus time. (b) Horizontal position of the droplet (black) and flame chemiluminescence (red) centroids subtracted by their corresponding mean values versus time. (c--j) are the flame chemiluminescence (top panels) and droplet shadowgraphy images (bottom panels) corresponding to $t = 0$, 0.08, 0.13, 0.14, 0.22, 0.46, 0.65, and 0.80~s, respectively. The results are for the representative test condition of BE0.01.}
\label{fig:B60E40Regimes}
\end{figure}

Compared to the biodiesel droplets, significantly different dynamics are observed for ethanol blended with biodiesel droplets. Similar dynamics are observed for all doping concentrations of ethanol blended with biodiesel, and the representative test condition of BE0.01 is selected and discussed here. The variations of the droplet diameter squared as well as the normalized flame chemiluminescence are shown by the black and red curves in Fig.~\ref{fig:B60E40Regimes}(a). The droplet and flame chemiluminescence centroids lateral position fluctuations are presented in Fig.~\ref{fig:B60E40Regimes}(b) by the black and red curves, respectively. Figures~\ref{fig:B60E40Regimes}(c--j) present the flame chemiluminescence (top panel) and shadowgraphy (bottom panel) images simultaneously acquired at $t = 0$, 0.08, 0.13, 0.14, 0.22, 0.46, 0.65, and 0.80~s, respectively. Also overlaid on the flame chemiluminescence images are the blue circles (representing the droplet with diameter $D$) and the flame chemiluminescence centroid, shown by the cross data symbol. As evident by the shadowgraphy image in Fig.~\ref{fig:B60E40Regimes}(c), almost immediately after the droplet is lit, nucleation occurs. This is due to the significantly smaller boiling temperature of ethanol (about 78$^\mathrm{o}$C) compared to that of biodiesel ($\gtrsim 200^\mathrm{o}\mathrm{C}$, from the safety data sheet provided by the biodiesel producer). The results in Fig.~\ref{fig:B60E40Regimes}(c--g) show that the boiling initially occurs due to homogeneous nucleation. As shown in Fig.~\ref{fig:B60E40Regimes}(a), both the droplet diameter squared and the normalized flame chemiluminescence exhibit pronounced fluctuations, which are due to early atomizations occurring immediately after the droplet is lit ($t\lesssim 0.3~\mathrm{s}$). Comparison of the results in Figs.~\ref{fig:B60E40Regimes}(b)~and~\ref{fig:B100Regimes}(b) show that, the addition of ethanol leads to large and random fluctuations of the flame lateral position and the mitigation of the coupling between this parameter and the lateral position of the droplet centroid. After the above intense and early atomization period, for $t \gtrsim 0.3~\mathrm{s}$, the heterogeneous nucleation at the liquid-solid interfaces occurs, leading to relatively mild atomizations, see Figs.~\ref{fig:B60E40Regimes}(h--j). For ethanol blended with biodiesel droplets, compared to biodiesel droplets, the atomizations cause several local extinctions, but the flame does not globally extinct, see for example Figs.~\ref{fig:B60E40Regimes}(e)~and~(h). \par

\begin{figure}[h]
	\centering
	\includegraphics[width=1\textwidth]{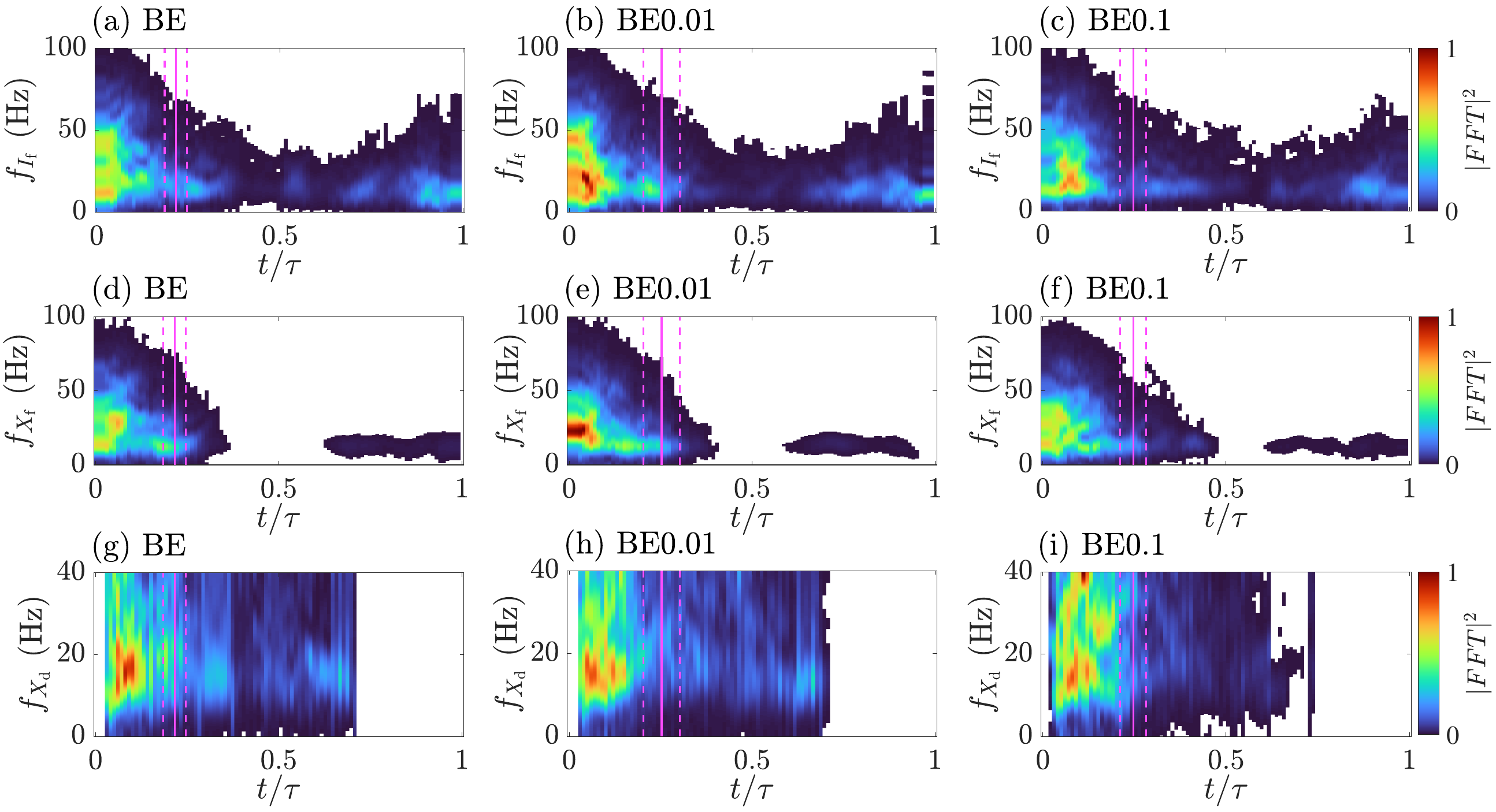}
	\caption{The spectrograms of ethanol blended with doped biodiesel single droplet combustion. (a--c) are for spatially integrated flame chemiluminescence, (d--f) are for the horizontal location of the flame chemiluminescence centroid, and (g--i) are for the horizontal location of the droplet center. The results in the first, second, and third columns pertain to ethanol blended with biodiesel, ethanol blended with biodiesel and doped with 0.01\% by mass of GO, and ethanol blended with biodiesel and doped with 0.1\% by mass of GO, respectively. The solid magenta line is the normalized time at which the homogeneous atomizations are completed.}
	\label{fig:BEsingle}
\end{figure}

The spectrograms of the flame chemiluminescence, the flame chemiluminescence centroid, and the lateral position of the droplet centroid were obtained and the results are presented in Fig.~\ref{fig:BEsingle}(a--c), (d--f), and (g--i), respectively. The results in the first, second, and third columns of Fig.~\ref{fig:BEsingle} pertain to ethanol blended with biodiesel droplets that are doped with 0, 0.01\%, and 0.1\% by mass of GO, respectively. As the results in Fig.~\ref{fig:BEsingle} show, the addition of graphene oxide to ethanol blended with biodiesel droplets does not substantially influence the dynamics of $I_\mathrm{f}$, $X_\mathrm{f}$, and $X_\mathrm{d}$. However, comparison of the results presented in Fig.~\ref{fig:Bsingle} with those in Fig.~\ref{fig:BEsingle} suggests differences. As can be seen, as soon as the ethanol containing droplets are lit, large amplitude oscillations for $f \lesssim 50$~Hz appear which are due to the early and intense homogeneous atomizations shown in Figs.~\ref{fig:B60E40Regimes}(c and d). The mean normalized droplet lifetime at which such homogeneous atomizations are completed is shown by the solid straight magenta line. The dashed lines present the above normalized time plus and minus half of the standard deviation of the normalized time, estimated based on the 12 repeats of each test condition. The shadowgraphy images (not presented here) and the results in Fig.~\ref{fig:BEsingle} show that, after about 50\% of the droplet lifetime, the intense atomizations are followed by heterogeneous atomizations, with a frequency of about 10~Hz.

\begin{figure}[h]
	\centering
	\includegraphics[width=1\textwidth]{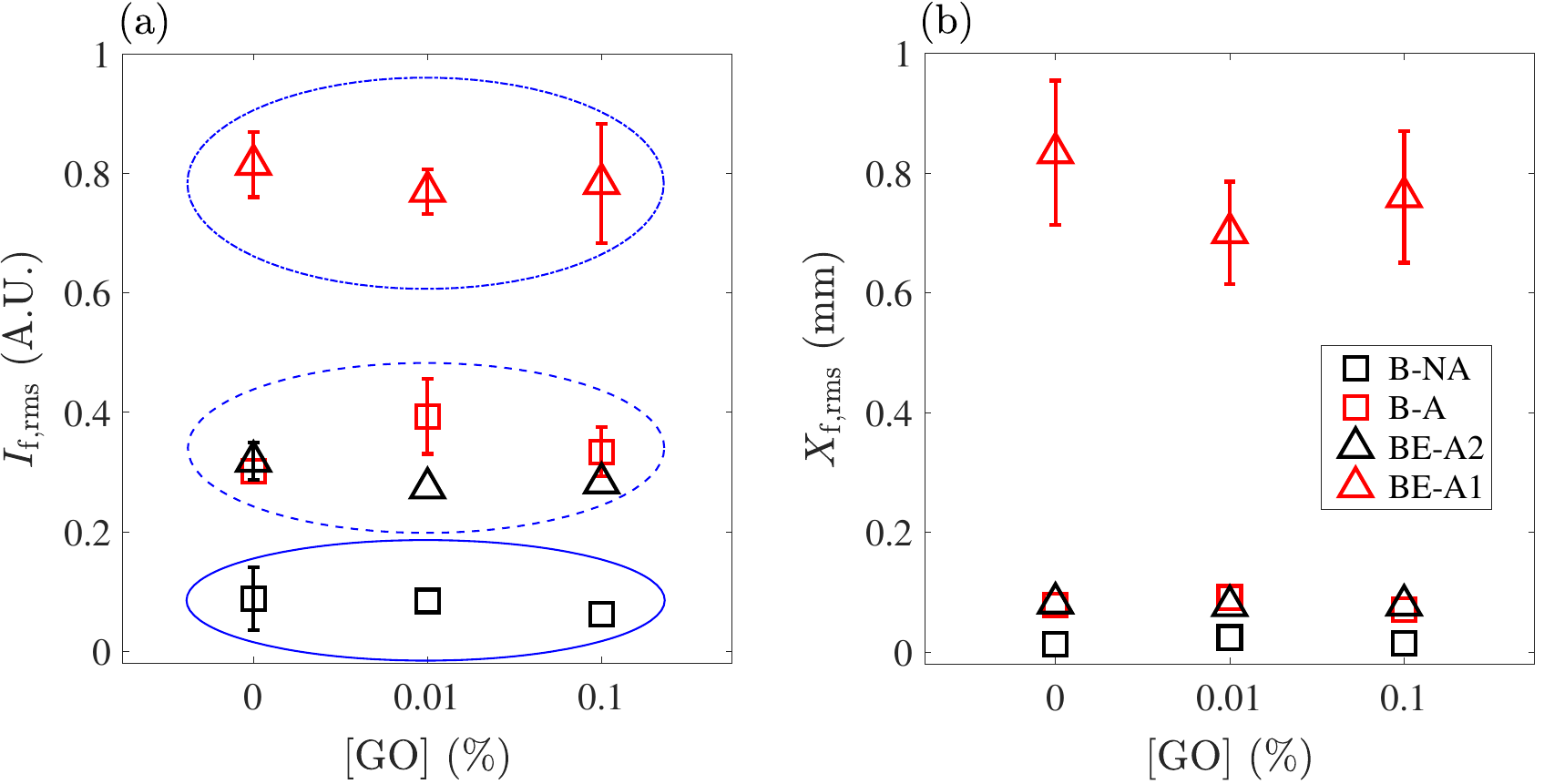}
	\caption{(a) and (b) are the rms of the flame chemiluminescence and the rms of the flame chemiluminescence centroid horizontal position, respectively. In the figure, B-NA and B-A correspond to before and after the heterogeneous atomizations, respectively, and for test conditions for which GO-doped biodiesel was used as the fuel. BE-A1 and BE-A2 correspond to the ethanol blended with biodiesel droplets that featured homogeneous and heterogeneous atomizations, respectively. The solid, dashed, and dotted-dashed ellipses correspond to test conditions with no atomizations, heterogeneous atomizations, and homogeneous atomizations, respectively.}
	\label{fig:IandXrms}
\end{figure}

The results presented in Figs.~\ref{fig:B100Regimes}--\ref{fig:BEsingle} suggest, though the addition of GO may mildly alter the biodiesel single droplet combustion dynamics, substantial alterations to these dynamics are observed as a result of blending with ethanol. Aiming to quantify such alterations, the rms of $I_\mathrm{f}$ and $X_\mathrm{f}$ were calculated. Since $X_\mathrm{d}$ cannot be obtained towards the end of the droplet lifetime, the rms of this parameter was not used for comparisons. In calculating the rms of $I_\mathrm{f}$ and $X_\mathrm{f}$, the corresponding data was partitioned into two segments, corresponding to periods of intense and moderate atomizations, which are related to homogeneous and heterogeneous atomizations, respectively. To this end, the normalized times highlighted by the magenta color solid lines in Figs.~\ref{fig:Bsingle}~and~\ref{fig:BEsingle} were used. Specifically, for biodiesel droplets, the root-mean-squares of $I_\mathrm{f}$ and $X_\mathrm{f}$ were calculated for time periods before (B-NA) and after (B-A) the atomizations. For ethanol blended with biodiesel droplets, the rms of the above data ($I_\mathrm{f,rms}$ and $X_\mathrm{f,rms}$) was calculated for those associated with the homogeneous atomizations (BE-A1) and those corresponding to after this period (BE-A2). The rms of $I_\mathrm{f}$ and $X_\mathrm{f}$ versus GO concentration are presented in Figs.~\ref{fig:IandXrms}(a)~and~(b), respectively. The error bars in the figures are the standard deviation of the corresponding data and are calculated based on the repeats of each tested fuel. Data points without an error bar featured standard deviations smaller than the size of the data symbol. The results in Fig.~\ref{fig:IandXrms} show that, generally, the addition of graphene oxide does not substantially influence the rms of $I_\mathrm{f}$ and $X_\mathrm{f}$. For biodiesel droplets, $I_\mathrm{f,rms}$ and $X_\mathrm{f,rms}$ are relatively large for the periods that the heterogeneous atomizations occur compared to those for no atomization periods (compare the black and red square data points). For ethanol blended with biodiesel droplets, $I_\mathrm{f,rms}$ and $X_\mathrm{f,rms}$ during the homogeneous atomizations are at least twice those during the heterogeneous atomizations (compare the red with black triangular data points). Also, the results in Fig.~\ref{fig:IandXrms} show that $I_\mathrm{f,rms}$ and $X_\mathrm{f,rms}$ are similar during the heterogeneous atomizations for both biodiesel and ethanol blended with biodiesel droplets (compare the red square with the black triangular data points). In summary, the rms of the flame chemiluminescence data that did not feature atomizations is grouped by the solid ellipse; and, that for data that featured heterogeneous and homogeneous atomizations are grouped by the dashed and dotted-dashed ellipses, respectively, in Fig.~\ref{fig:IandXrms}(a). Similar groups are also identified for the results presented in Fig.~\ref{fig:IandXrms}(b) but the ellipses are not shown for clarity of the presentation. Overall, the results show that, compared to the heterogeneous atomizations, the homogeneous atomizations lead to at least 2 and 7 times larger values of $I_\mathrm{f,rms}$ and $X_\mathrm{f,rms}$, respectively. This finding has implications for the dynamics of multi-droplets combustion as well as how fast the flame spreads in a multi-droplet configuration. The former and the latter are studied in the next subsection and next section, respectively.

\subsubsection{Multi-droplet combustion dynamics}
\label{subsec:multidynamics}
\begin{figure}[h]
	\centering
	\includegraphics[width=1\textwidth]{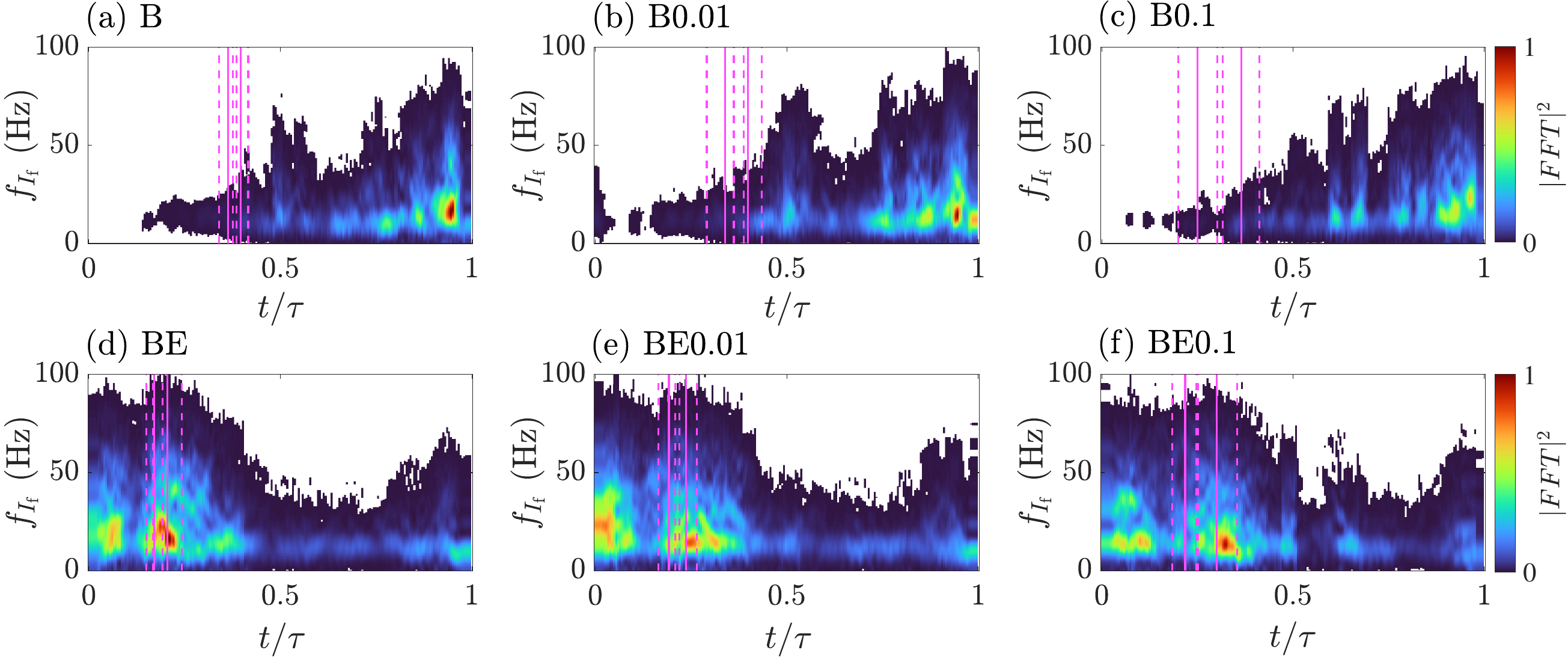}
	\caption{The spectrograms of the flame chemiluminescence for multi-droplet combustion with the center droplet lit. The results in (a--c) correspond to biodiesel and (d--f) correspond to ethanol blended with biodiesel. The results in the first, second, and third columns pertain to neat fuel, fuel doped with 0.01\% by mass of GO, and 0.1\% by mass of GO, respectively. The magenta solid lines show the mean normalized times at which the the side droplets are enveloped by the flame.}
	\label{fig:multidropletscenter}
\end{figure}

\begin{figure}[h]
	\centering
	\includegraphics[width=1\textwidth]{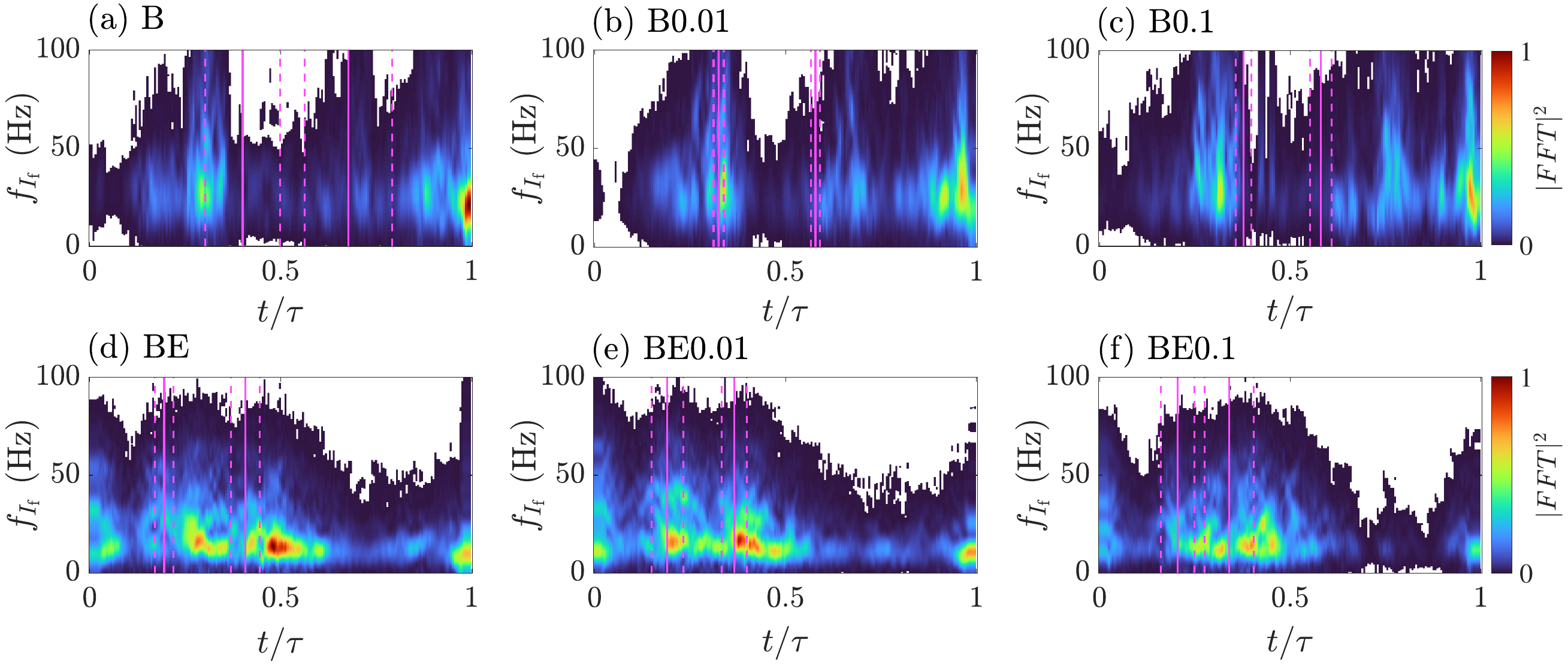}
	\caption{The spectrograms of the flame chemiluminescence for multi-droplet combustion with the rightmost droplet lit. The results in (a--c) correspond to biodiesel and (d--f) correspond to ethanol blended with biodiesel. The results in the first, second, and third columns correspond to neat fuel, fuel doped with 0.01\% by mass of GO, and 0.1\% by mass of GO, respectively. The magenta solid lines present the mean normalized times at which the side droplets are enveloped by the flame.}
	\label{fig:multidropletsside}
\end{figure}

Once three droplets were deposited on the suspending fibers, either the droplet at the center or that on the right-hand-side was lit, with the lit droplet referred to as the main droplet. The droplet to which the propagation occurs is referred to as the side droplet. For spectral analysis, the horizontal position of the droplet centroid is not used here, since more than one droplet is present and analysis similar to those presented for single droplet combustion could not be performed. Also, the horizontal location of the flame chemiluminescence centroid did not allow for best presentation of the signature of the atomization events and was not used here. Instead, the spectrogram of the spatially integrated flame chemiluminescence was used to study the spectral characteristics of the multi-droplet combustion. Such spectral characteristics depend on whether the droplet at the center or that on the side is lit. Thus, the results corresponding to the former and the letter are grouped accordingly and presented in Figs.~\ref{fig:multidropletscenter}~and~\ref{fig:multidropletsside}, respectively. The results in the first and second rows of the figures correspond to biodiesel and ethanol blended with biodiesel, respectively. Those in the first, second, and third columns of both figures correspond to the neat fuel, fuel doped with 0.01\% by mass of GO, and fuel doped with 0.1\% by mass of GO, respectively. Overlaid on both figures are the straight solid lines in magenta color which present the averaged normalized time at which the flame envelopes the side droplets. The normalization is performed with respect to the three-droplet lifetime. The dashed lines in magenta color present the above averaged and normalized times plus and minus half of a standard deviation. Both the averaging and standard deviation calculations are performed based on the 12 repeats of the experiments for each tested fuel.

The temporal separation between the solid magenta lines in Figs.~\ref{fig:multidropletscenter}~and~\ref{fig:multidropletsside} allows to understand the time difference between the flame enveloping the side droplets. For test conditions that the center droplet is lit, the above separation time is smaller than that for the test conditions that the farthest (right-hand-side) droplet is lit, compare the distance between the magenta solid lines in Fig.~\ref{fig:multidropletscenter} with those in Fig.~\ref{fig:multidropletsside}. This difference is because once the center droplet is lit, heat transfer to both side droplets start. However, for scenarios that the farthest droplet is lit first, the last droplet is lit only after the center droplet is ignited, leading to long delay for the ignition of the farthest droplet. For biodiesel test conditions with the center droplet lit, the ignition of both side droplets occur at a relatively short time period from one another, and the heterogeneous atomizations (which occur towards the end of the multi-droplet combustion lifetime, see Figs.~\ref{fig:multidropletscenter}(a--c)) occur after the propagation to the side droplets are completed. Nonetheless, the heterogeneous atomizations are accompanied by the large amplitude oscillations near $\sim$10~Hz, which is similar to that observed for single biodiesel droplet combustion. For biodiesel droplets that the farthest droplet is lit, heterogeneous atomizations occur at the end of the side droplets (center and farthest droplets) lifetime (see the large amplitude oscillations at $t/\tau \approx 0.3$ in Fig.~\ref{fig:multidropletsside}(a--c)) which is followed by the ignition of the farthest droplet. Towards the end of the farthest droplet combustion lifetime, the heterogeneous atomizations occur, see the large oscillations amplitude in the second half of the multi-droplet lifetime in Figs.~\ref{fig:multidropletsside}(a--c).

For multi-droplet combustion of ethanol blended with biodiesel and doped with GO, the results presented in the second rows of Figs.~\ref{fig:multidropletscenter}~and~\ref{fig:multidropletsside} show relatively large values of the spectrograms near $t/\tau \approx 0.1$, which was confirmed to be due to the occurrence of early and intense homogeneous atomizations similar to the results presented in Fig.~\ref{fig:B60E40Regimes} for single droplet combustion of ethanol blended with biodiesel and doped with GO. For multi-droplet combustion of the above fuel, the early homogeneous atomizations are followed by the early ignition of the second droplet, compare the normalized times associated with the first magenta solid lines in the second rows of Figs.~\ref{fig:multidropletscenter}~and~\ref{fig:multidropletsside} with those in the first rows of the figures. It is also observed that, for either the center or rightmost droplet ignition scenarios, the last droplet ignited sooner for the ethanol blended with biodiesel test conditions compared to biodiesel test conditions, note the time difference between the magenta solid lines in the first and second rows of Figs.~\ref{fig:multidropletscenter}~and~\ref{fig:multidropletsside} and for matching tested fuels. Finally, the results in Figs.~\ref{fig:multidropletscenter}(d--f)~and~\ref{fig:multidropletsside}(d--f) show that after the flame propagation to the last droplet is completed, the ethanol blended with biodiesel and doped with GO droplets continue to feature atomizations and towards the end of the multi-droplet lifetime. The shadowgraphy videos showed that the large values of the FFT immediately after the second solid magenta lines are associated with heterogeneous atomizations.

\begin{figure}[h]
	\centering
	\includegraphics[width=1\textwidth]{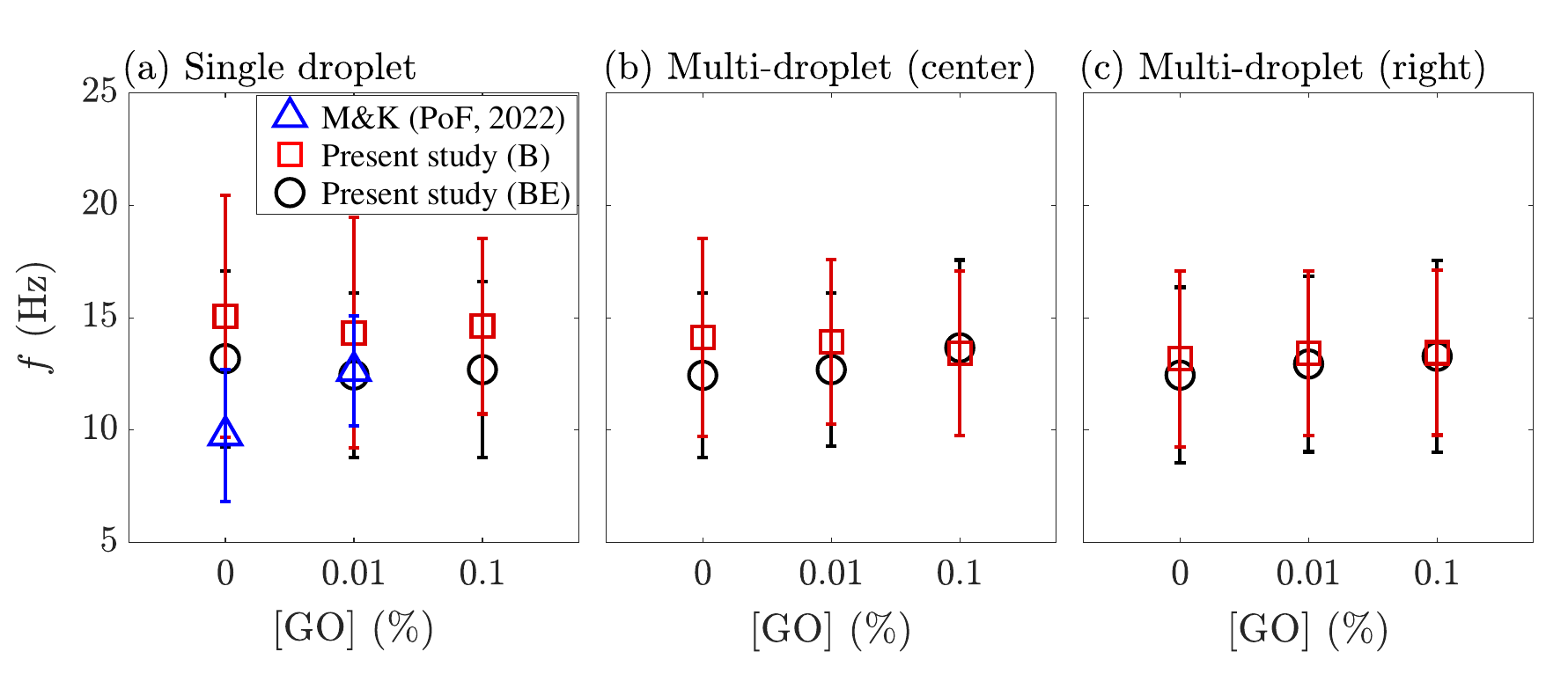}
	\caption{The dominant frequency of the flame chemiluminescence oscillations. The results in (a), (b), and (c) correspond to single droplet combustion, multi-droplet combustion with the center droplet ignited, and multi-droplet combustion with the rightmost droplet ignited. The triangular, square, and circular data points are for diesel (extracted from \cite{mosadegh2022role}), biodiesel, and ethanol blended with biodiesel, respectively.}
	\label{fig:multifrequency}
\end{figure}

To compare the spectral contents of the dynamics presented in Figs.~\ref{fig:multidropletscenter}~and~\ref{fig:multidropletsside}, the FFT magnitude of the flame chemiluminescence oscillations are calculated during the droplets lifetime, the dominant frequency of oscillations are obtained, and the results are presented in Fig.~\ref{fig:multifrequency}. The length of the error bar is the full-width at half maximum of the FFT magnitude at the most dominant frequency. The results in Figs.~\ref{fig:multifrequency}(a--c) correspond to single droplet, multi-droplet with center droplet lit, and multi-droplet with the rightmost droplet lit, respectively. For comparison purposes, the results of single droplet combustion from Mosadegh and Kheirkhah~\cite{mosadegh2022role} (performed for diesel) are also extracted and presented in Fig.~\ref{fig:multifrequency}(a) with the blue triangular data points. The results show that, overall, the small-frequency oscillations observed for single droplet combustion also exists for multi-droplet combustion, with the atomizations being the root-cause of the oscillations. It is also observed that the addition of neither ethanol nor GO substantially influences the above frequency. Whether such small-frequency oscillations also exists for a multi-droplet configuration with more than three droplets or for turbulent spray flames is not studied in this work and is a topic for future investigations.

\subsection{Effect of atomization on the rate of flame spread among multi-droplets}

\begin{figure}[h]
\centering
\includegraphics[width=1\textwidth]{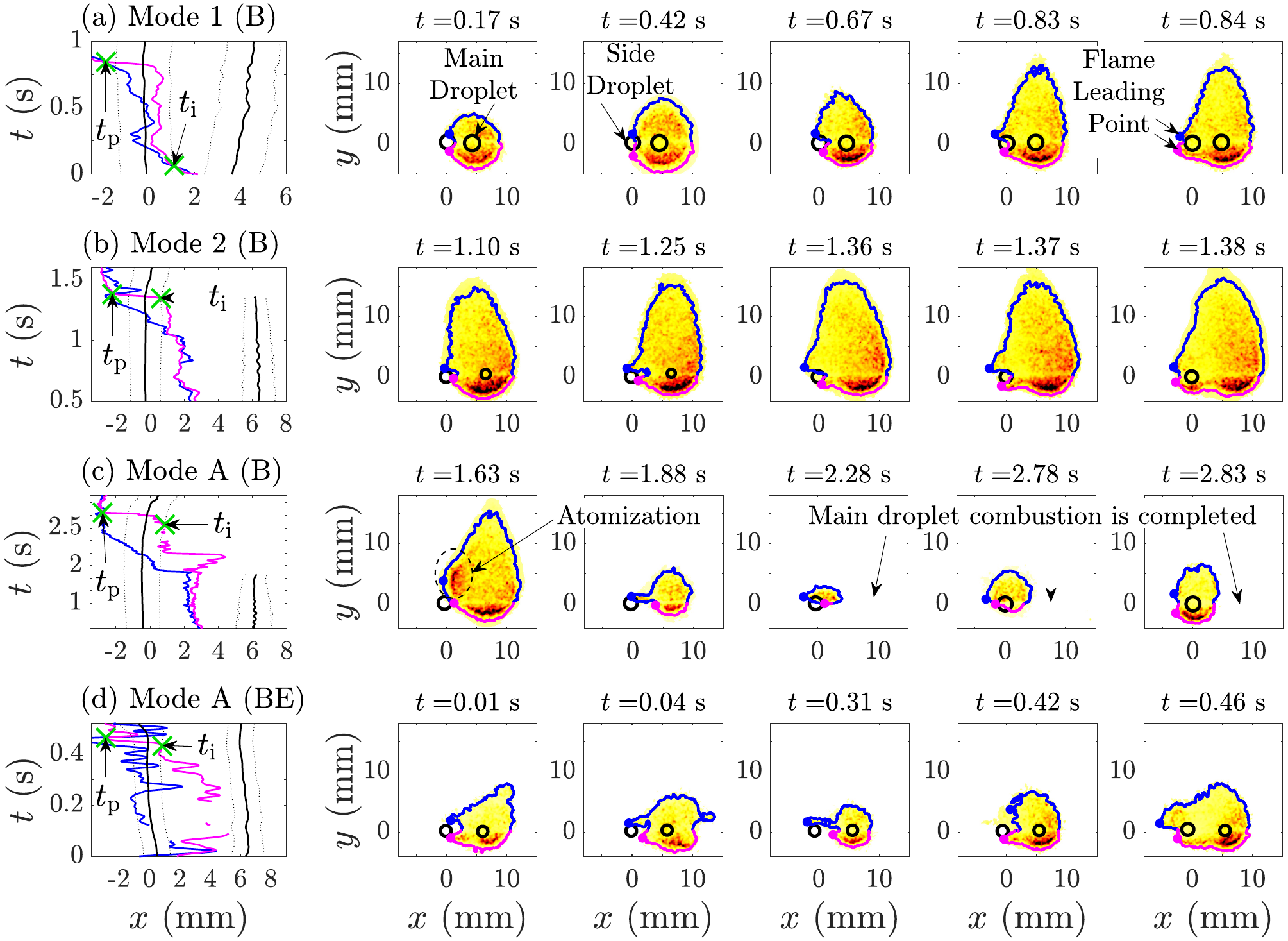}
\caption{The identified modes of flame spread. The first and second rows correspond to Modes~1 and 2 of flame propagation. The results in the third and fourth rows correspond to modes of flame spread due to heterogeneous and homogeneous atomizations, respectively. The results presented in (a--c) and (d) correspond to B0.1 and BE0.1, respectively. The results in the first column demonstrate the ignition and spread times and how they were obtained. The videos for the results presented in the first to fourth rows are provided as supplementary materials V1--V4, respectively.}
\label{fig:modes}
\end{figure}

The flame spread rate is substantially influenced by the mode of the flame spread~\cite{mikami2006microgravity}. In total, four modes of flame spread were identified, with characteristics of propagation and five chemiluminescence images (for each mode) presented in each row of Fig.~\ref{fig:modes}. Overlaid on Fig.~\ref{fig:modes} are the main and side droplets shown by the black circular data points. The size of the circular data points resemble the diameter of the relevant droplet. For the figures that the main droplet is not shown, its size is close to the suspending fibers and the droplet cannot be detected from the shadowgraphy images. Additionally, videos relevant to the results in Fig.~\ref{fig:modes} are also provided as supplementary materials V1--V4. It is important to note that, while the dynamics of spread between two droplets are shown in Fig.~\ref{fig:modes}, the third droplet may or may not be lit. For example, for the results shown in supplementary material V1, the rightmost droplet was initially lit and video V1 presents the flame propagation that takes place between the center and leftmost droplet. It is observed that for all tested fuels, the flame propagates at the top before it propagates at the bottom. Since the experiments were conducted under normal gravity conditions, the upward natural convection increases the heat transfer to the upper portion of the side droplet and this is speculated to increase the fuel vapor formed at the top of the side droplet and as a result the sooner propagation of the flame at the top. Following the discussions in the supplementary materials, the flame boundaries at the top and the bottom of the main droplet are obtained and overlaid by the solid blue and purple curves on the flame chemiluminescence images in Fig.~\ref{fig:modes}. Also, the leading points of the flame both at the top and at the bottom (the points on the most left position of the blue and purple curves) were monitored during time. The horizontal locations of the flame's leading point at the top (bottom) is shown using the blue (purple) color in Figs.~\ref{fig:modes}(a--d) (see the first column of the figure). In these figures, the right (left) solid black curve presents the horizontal location of the main (side) droplet center versus time. In Figs.~\ref{fig:modes}(a--d), for convenience, the horizontal axis presents the horizontal position; and, the vertical axis is time. In the figures, the dotted black curves present the horizontal locations of the droplets edges. The time instant at which the leading point of the bottom flame contour reaches the side droplet is referred to as $t_\mathrm{i}$; and, $t_\mathrm{p}$ is the time instant that the horizontal locations of the leading points at the top and bottom coincide. Both $t_\mathrm{i}$ and $t_\mathrm{p}$ are highlighted by the green cross data symbol on the $x-t$ diagrams in Figs.~\ref{fig:modes}(a--d). In the present study, the ignition time ($\tau_\mathrm{i}$, i.e. the time a flame kernel is formed at the bottom of the side droplet) is measured from the moment a flame is formed around the main droplet and as such $\tau_\mathrm{i} = t_\mathrm{i}$. The propagation time ($\tau_\mathrm{p}$) is the time it takes for the horizontal locations of the leading points at the top and bottom of the side droplet to coincide with respect to the ignition time of the side droplet, i.e. $\tau_\mathrm{p} = t_\mathrm{p} - t_\mathrm{i}$.

Analysis of the flame chemiluminescence images showed that the presence of ethanol as well as the mean droplet spacing significantly influence the modes of flame spread. The latter is denoted by $\overline{S}$ here. This parameter is the distance between the main and side droplets averaged between $t=0$ and the smaller between $t_\mathrm{i}$ and the latest time that the main droplet could be detected from the shadowgraphy images. For all test conditions, $\overline{S}$ varied from about $1.7D_0$ to $3.7D_0$. It was observed that GO-doped biodiesel droplets with $1.7 \lesssim \overline{S}/D_0 \lesssim 2.2$ featured Mode~1 of flame spread, which is similar to the observation in Mikami~\textit{et al.}~\cite{mikami2006microgravity}. For this mode, as can be seen by the exemplary flame chemiluminescence images, the flame spread is completed before the occurrence of the atomization events, and as a result, the flame spread is not affected by the atomizations. Mode~2 of the flame spread occurs for the values of $\overline{S}$ that are large enough that the flame of the main droplet does not reach the side droplet immediately after the flame of the main droplet is formed. The flame chemiluminescence images for Mode~2 (see the flame chemiluminescence images in the second row of Fig.~\ref{fig:modes}) show that during about 1~s, the flame of the main droplet is established, and the side droplet receives heat from this flame. This is speculated to be followed by formation of an ignitable fuel vapor and air mixture between the side droplet and the main droplet flame, and as a result, the flame of the main droplet propagates towards the side droplet. It was observed that Mode~2 occurs for $2.2 \lesssim \overline{S}/D_0 \lesssim 3.5$, which agrees with the findings of Mikami~\textit{et al.}~\cite{mikami2006microgravity}. \par

The third and fourth modes of flame spread, identified in the present study, are associated with the heterogeneous and homogeneous atomizations, respectively. For GO-doped biodiesel droplets, provided the inter-droplet spacing is large enough ($\overline{S}/D_0 \gtrsim 3.5$), the flame of the main droplet does not propagate towards the side droplet initially. However, towards the end of the main droplet lifetime, this droplet undergoes heterogeneous atomizations and due to the flame's lateral movement, the main droplet flame eventually reaches the side droplet and propagates around this droplet, see the flame chemiluminescence image corresponding to $t = 1.64$~s in the third row of Fig.~\ref{fig:modes}. This mode of propagation is referred to as Mode~A~(B), with ``A" and ``B" standing for atomizations and biodiesel, respectively. For this mode, typically, the main droplet combustion is completed, and the kernel formed around the side droplet slowly propagates to envelop the side droplet, see the flame chemiluminescence images presented in the third row of Fig.~\ref{fig:modes}. Finally, for ethanol blended droplets and for $1.7 \lesssim \overline{S}/D_0 \lesssim 3.7$, due to homogeneous atomizations, the side droplet is ignited immediately after the main droplet flame is formed. Compared to the biodiesel multi-droplet combustion for which the main droplet is completely burnt before the side droplet flame fully envelopes this droplet, for ethanol blended with biodiesel multi-droplets, the main flame envelops the side droplet and during the homogeneous atomizations. This last mode of flame propagation is referred to as Mode~A~(BE), with ``BE" standing for ethanol blended with biodiesel. \par

%For Mode 1, as shown by the exemplary results in Fig.~\ref{fig:modes}(a), during a relatively short period of time, the side droplet is ignited (here $t_\mathrm{i}=0.17~\mathrm{s}$). Then, as shown in the first row of Fig.~\ref{fig:modes}, the flame initially propagates along the upper part of the side droplet before it eventually propagates at the base of the side droplet.

\begin{figure}[h]
\centering
\includegraphics[width=1\textwidth]{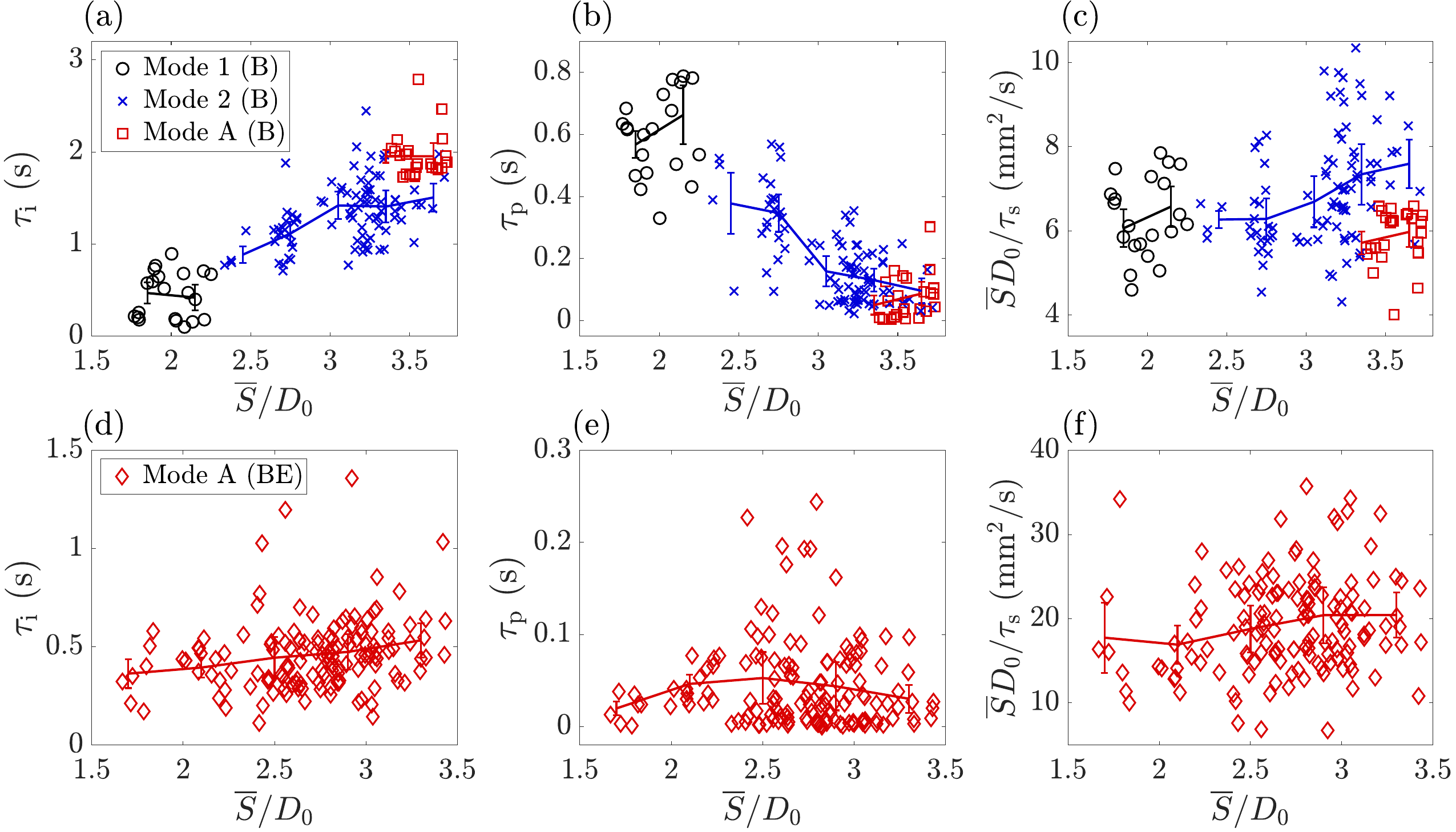}
\caption{Ignition time, propagation time, and flame spread rate versus the normalized mean inter-droplet spacing. (a--c) and (d--f) correspond to the biodiesel and ethanol blended with biodiesel multi-droplet combustion, respectively.}
\label{fig:spread}
\end{figure}

For all of the above identified modes, the ignition, propagation, and flame spread times (the time that it takes for the flame to envelop the side droplet after the main droplet is lit, i.e. $\tau_\mathrm{s} = \tau_\mathrm{i}+\tau_\mathrm{p}$) were calculated. For all test conditions, the side droplet ignition time, the flame propagation time around the side droplet, and the flame spread rate were obtained and presented against the normalized spacing in the first, second, and third columns of Fig.~\ref{fig:spread}, respectively. The results pertaining to biodiesel and ethanol blended with biodiesel droplets are shown in Figs.~\ref{fig:spread}(a--c) and Figs.~\ref{fig:spread}(d--f), respectively. Data presenting Modes~1~and~2 are shown by the black circles and blue cross symbols, respectively. Data relating to Mode A~(B) is presented using the red square symbol, and that for Mode A~(BE) is shown using the red diamond symbol. The scatter in the data in each panel is due to the 12 repeats of each test condition, the three tested concentrations of GO, two tested configurations for multi-droplet combustion (which are lighting the side or center droplets), and finally two propagation scenarios for each configuration (right to center and center to left droplets as well as center to two side droplets). Thus, $12 \times 3 \times 2 \times 2  = 144$ data points exist in each panel of Fig.~\ref{fig:spread}. To study the impact of spacing on $\tau_\mathrm{i}$, $\tau_\mathrm{p}$, and $\overline{S}D_0/\tau_\mathrm{s}$ as well as for later comparisons between the results of the present study and those of others, the scattered data points in Fig.~\ref{fig:spread} were grouped into several bins; and, the data in the bins was averaged. Finally, for each mode of the flame spread, the averaged data points were presented by the solid lines in Fig.~\ref{fig:spread}. These lines facilitate understanding the influences of inter-droplet spacing and modes of flame spread on $\tau_\mathrm{i}$, $\tau_\mathrm{p}$, and $\overline{S}D_0/\tau_\mathrm{s}$. In Fig.~\ref{fig:spread}, the length of the error bars are the standard deviation of the scattered data point within each bin. \par 

The results in Fig.~\ref{fig:spread}(a) show that, for Mode~1, $\tau_\mathrm{i}$ is nearly constant. However, for Mode~2, $\tau_\mathrm{i}$ is relatively larger than that for Mode~1 and the ignition time increases with increasing the normalized spacing. This is because, with increasing $\overline{S}/D_0$, the transfer of heat from the flame of the main droplet to the side droplet is decelerated, and as a result, $\tau_\mathrm{i}$ increases. The results in Fig.~\ref{fig:spread}(a) show that, for $\overline{S}/D_0 \gtrsim 3.5$, both Modes~2~and~A~(B) may co-exist. It is observed that, for a fixed value of the inter-droplet spacing, $\tau_\mathrm{i}$ is larger for Mode~A~(B) than that for Mode~2. This suggests that, for Mode~A~(B), the heterogeneous atomizations distribute the heat generated from the main droplet combustion in random directions leading to decreased rate of heat transfer to the side droplet and as a result larger values of $\tau_\mathrm{i}$ compared to those for Mode~2. Nonetheless, for the large values of $\overline{S}/D_0$ examined in this study, atomizations can lead to the ignition and combustion of the side droplet via Mode A~(B) and in addition to Mode~2.

Compared to $\tau_\mathrm{i}$ that depends on how soon a flame kernel is formed at the vicinity of the side droplet, $\tau_\mathrm{p}$ is influenced by the amount of the combustible mixture that is formed around the side droplet. The results presented in Figs.~\ref{fig:spread}(b) show that $\tau_\mathrm{p}$ is nearly scattered for the tested normalized droplets spacing and for Mode~1. However, for Modes~2 and A~(B), increasing the inter-droplet spacing decreases $\tau_\mathrm{p}$. This is because the relatively large values of $\tau_\mathrm{i}$ for Modes~2 and A~(B) compared to Mode~1 facilitate longer periods of heat transfer and as a result formation of relatively larger amount of fuel vapour around the side droplet, which leads to faster flame propagation (or smaller values of $\tau_\mathrm{p}$ for Modes~2 and A~(B) compared to those for Mode~1). \par

Compared to biodiesel droplets combustion, for ethanol blended with biodiesel droplets combustion, the main droplet atomizations begin as soon as it is lit, which substantially influences how fast the flame spreads. Comparison of the results presented in Fig.~\ref{fig:spread}(a) with those in Fig.~\ref{fig:spread}(d) shows that, the ignition time for Mode A~(BE) is significantly smaller than that for Mode A~(B). This is due to the intense homogeneous atomizations for Mode~A~(BE) compared to relatively mild heterogeneous atomizations of Mode~A~(B). The intense homogeneous atomizations increase the possibility of ignition and leads to smaller values of $\tau_\mathrm{i}$ for Mode~A~(BE). For this mode, the results presented in Fig.~\ref{fig:spread}(e) show that increasing $\overline{S}/D_0$ from about 1.7 to 2.5 increases the propagation time; however, further increasing the normalized spacing decreases $\tau_\mathrm{p}$. This is because, on one hand, increasing $\overline{S}/D_0$ increases the time for the droplets evaporation which can increase the amount of vapor around the droplet and decrease $\tau_\mathrm{p}$. However, on the other hand, increasing the inter-droplet spacing leads to significant loss of combustion heat from the droplet prior to the flame propagation, which could reduce the amount of fuel vapor around the side droplet and potentially decreasing $\tau_\mathrm{p}$. \par

The trends of variations for $\tau_\mathrm{i}$ and $\tau_\mathrm{p}$ presented in Figs.~\ref{fig:spread}(a, b, d, and e) influence the flames spread rate presented in Figs.~\ref{fig:spread}(c and e). It can be seen that, for Modes~1 and 2, the flame spread rate increases from about 6 to 8 increasing the normalized inter-droplet spacing from about 2 to 3. The reason for this increasing trend is due to the significant decrease of $\tau_\mathrm{p}$ by increasing the normalized spacing. Compared to those for Modes~1~and~2, relatively small values of $\overline{S}D_0/\tau_\mathrm{s}$ is reported for Mode~A~(B), which is due to the prolonged ignition time caused by atomizations. Compared to the results in Fig.~\ref{fig:spread}(c), those in Fig.~\ref{fig:spread}(f) show significant scatter of the flame spread rate (6--8~$\mathrm{mm^2/s}$ versus 10--40~$\mathrm{mm^2/s}$). Nonetheless, owing to relatively small values of both $\tau_\mathrm{i}$ and $\tau_\mathrm{p}$, values of the flame spread rate for droplets featuring homogeneous atomizations are approximately three times those featuring heterogeneous atomization. \par 

\begin{figure}[h]
	\centering
	\includegraphics[width=1\textwidth]{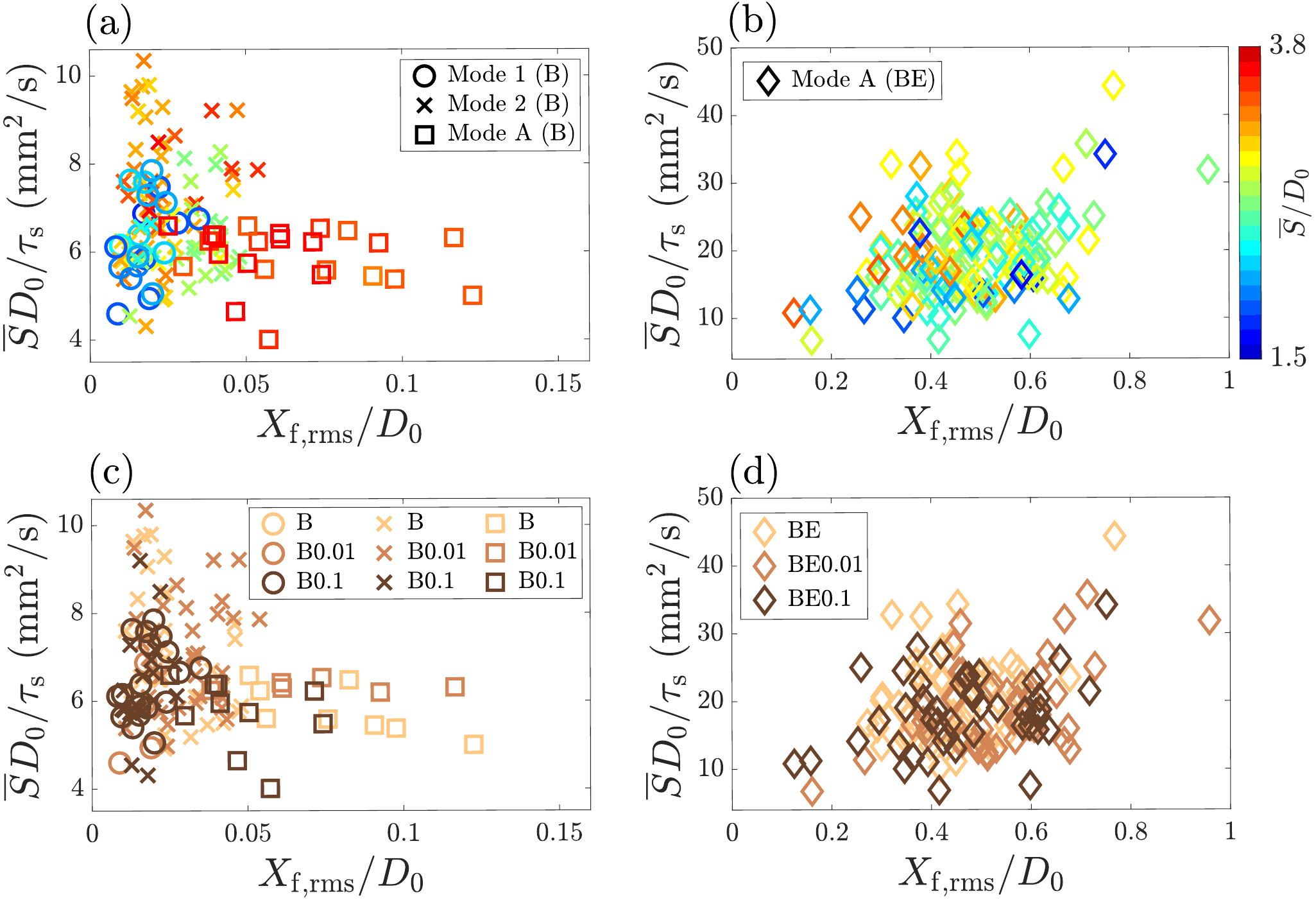}
	\caption{The flame spread rate versus the rms of the flame chemiluminescence centroid lateral movements during the spread time. (a and c) correspond to Modes~1,~2,~and~A~(B). (b and d) correspond to Mode A~(BE). Data in (a and b) are color-coded based on the inter-droplet spacing. Data in (c and d) are color coded based on GO doping concentration.}
	\label{fig:spreadrms}
\end{figure}

Though the results presented above and those in section~\ref{subsec:dynamics} suggested that blending ethanol with biodiesel is accompanied by significant increase of both the rms flame chemiluminescence centroid lateral position and the flame spread rate, the relation between the above parameters and how the inter-droplet spacing as well as the presence of GO influence the above relation are unknown. Aiming to investigate this, the rms of the flame chemiluminescence centroid lateral movements during the spread time ($X_\mathrm{f,rms}$) is calculated for all test conditions and its variation versus the flame spread rate is presented in Fig.~\ref{fig:spreadrms}. Please note that the results in Fig.~\ref{fig:spreadrms}(a) are identical to those in Fig.~\ref{fig:spreadrms}(c); and, those in Fig.~\ref{fig:spreadrms}(b) are identical to those in Fig.~\ref{fig:spreadrms}(d). For comparison purposes, however, the results are color-coded based on the GO concentration and the mean normalized inter-droplet spacing. The results in Figs.~\ref{fig:spreadrms}~(a and b) are color coded based on the normalized inter-droplet spacing; and, those in Figs.~\ref{fig:spreadrms}~(c and d) are color coded based on the GO doping concentration. The results in the first and second columns pertain to Modes~1,~2,~and A~(B) as well as A~(BE), respectively. It can be seen that the data points related to Modes~1 and 2 (see the circle and cross data symbols) are clustered, correspond to $X_\mathrm{f,rms}/D_0 \lesssim 0.05$, and feature flame spread rate smaller than 8~$\mathrm{mm^2/s}$. Comparison of the results for Modes~1 and 2 with those for Mode~A~(B) shows that, even though the heterogeneous atomizations increase $X_\mathrm{f,rms}$ for biodiesel, the flame spread rate remains unchanged and about 6~$\mathrm{mm^2/s}$. Compared to Modes~1,~2,~and A~(B), the results for Mode~A~(BE) show that the flame spread rate and $X_\mathrm{f,rms}$ are positively related. Specifically, increasing $X_\mathrm{f,rms}/D_0$ from 0.2 to 1 increases the flame spread rate from about 10 to 40~$\mathrm{mm^2/s}$. Comparison of the results in Figs.~\ref{fig:spread}(a) and (c) shows that, for biodiesel droplets, the largest tested doping concentration featured relatively small $X_\mathrm{f,rms}$ and flame spread rate. Moderate doping of biodiesel; however, leads to relatively large flame spread rate. Compared to biodiesel, the results in Figs.~\ref{fig:spread}(b and d) show that, the addition of GO does not influence the relation between the flame spread rate and $X_\mathrm{f,rms}$ for ethanol blended with biodiesel droplets. 

\begin{figure}[h]
\centering
\includegraphics[width=1\textwidth]{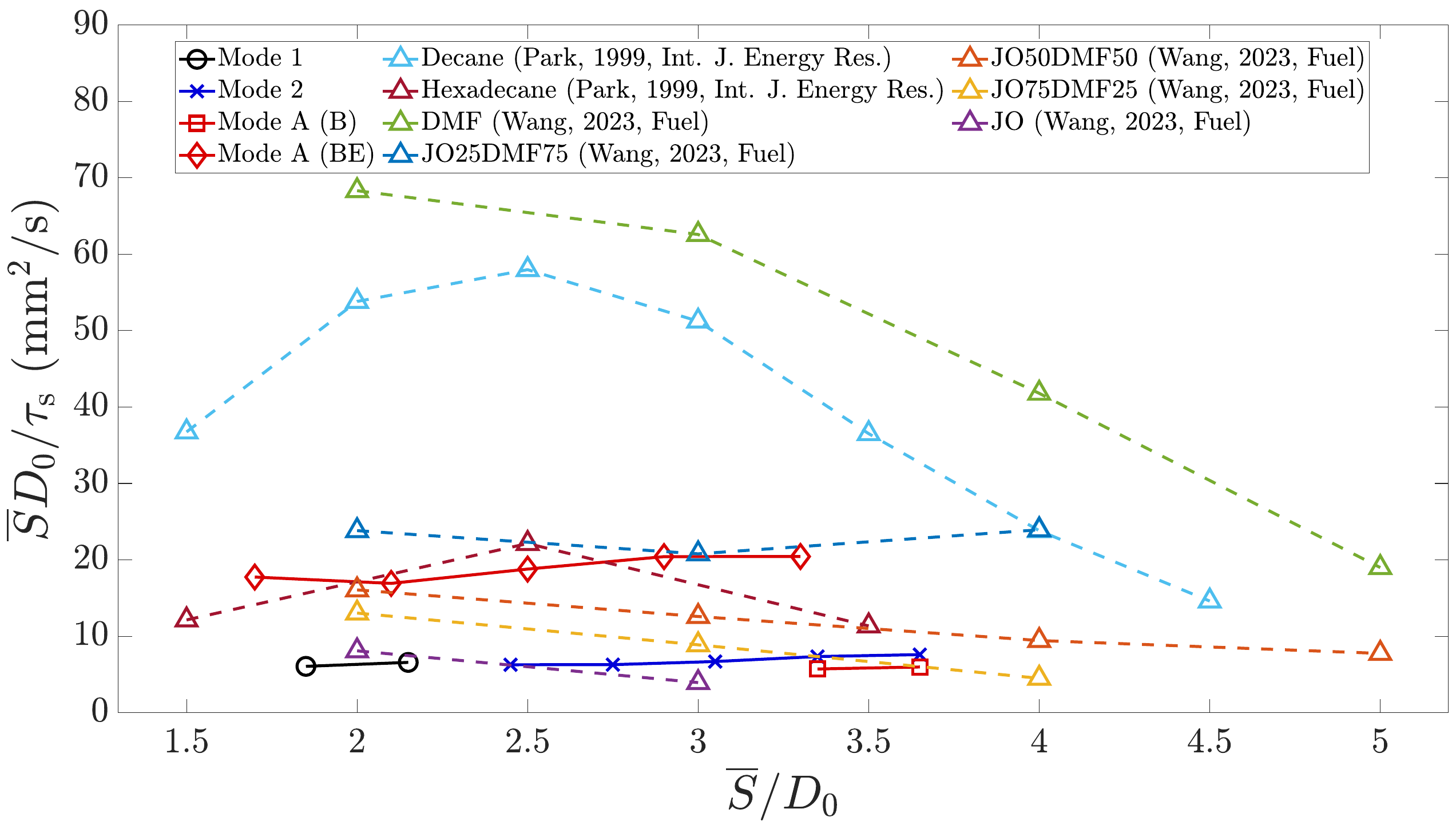}
\caption{The flame spread rate versus the normalized inter-droplet spacing. Overlaid on the figure are the results of Park~\textit{et al.}~\cite{park1999flame,park1999study} and Wang \textit{et al.}~\cite{wang2023micro}.}
\label{fig:spreadcomp}
\end{figure}

The results of the present study were compared against those in the literature. The segmented lines in Figs.~\ref{fig:spread}(c and f) versus the normalized spacing are presented in Fig.~\ref{fig:spreadcomp}. Overlaid on the figure are the results of Park~\textit{et al.}~\cite{park1999flame,park1999study} and Wang~\textit{et al.}~\cite{wang2023micro}. In the studies of Park~\textit{et al.}~\cite{park1999flame,park1999study}, the flame spread rate was measured for multi-droplets of decane and hexadecane. Wang \textit{et al.}~\cite{wang2023micro} measured the flame spread rate for multi-droplets of 5-dimethylfuran blended with Jatropha oil. The boiling temperatures of decane, DMF, hexadecane, and JO are reported as 174~$^{\mathrm{o}}$C~\cite{li2018composite}, 92~$^{\mathrm{o}}$C~\cite{wang2023micro}, 297~$^{\mathrm{o}}$C~\cite{park1999flame}, 355~$^{\mathrm{o}}$C~\cite{wang2023micro}, respectively. It can be inferred from the results in Fig.~\ref{fig:spreadcomp} that the small boiling temperature/large volatility of decane and DMF is accompanied by about an order of magnitude larger flame spread rate compared to those for Jatropha oil and biodiesel, which are fuels with relatively large boiling temperatures/low volatilities. The results of Wang \textit{et al.}~\cite{wang2023micro} also show that blending fuels with different volatilities leads to mixtures with flame spread rate values in between those of the original fuels. This agrees with the observations reported in the present study that the addition of ethanol (which is more volatile than biodiesel) increases the flame spread rate of biodiesel droplets. The results of both the present study and those of Park~\textit{et al.}~\cite{park1999flame,park1999study} and Wang~\textit{et al.}~\cite{wang2023micro} show that, for a given fuel blend, the flame spread rate is nearly insensitive to $\overline{S}/D_0$ and for the tested normalized inter-droplet spacing. However, for fuels with large volatility, the flame spread rate varies by changing the tested normalized inter-droplet spacing. In essence, the findings of past investigations and those in this work show that, for fuels with low volatility (such as biodiesel), the occurrence of homogeneous atomizations (compared to heterogeneous atomizations) significantly increases the flame spread rate.

\section{Concluding remarks}
\label{conclusion}

The effects of homogeneous and heterogeneous atomizations on both the combustion dynamics of multi-droplets and the flame spread rate among them were investigated experimentally. The utilized fuels were biodiesel and 60\% biodiesel blended with 40\% ethanol by mass. All fuels were doped with octadecyl amine functionalized graphene oxide with doping concentrations of 0, 0.01, and 0.1\% by mass. The suspensions of graphene oxide in biodiesel and ethanol blended with biodiesel were prepared, and it was confirmed that they were stable for time periods substantially longer than the duration of the combustion experiments. Simultaneous and temporally resolved flame chemiluminescence and shadowgraphy imaging were performed. Either one or three droplets were deposited at the intersections of silicon carbide fibers. A plasma-assisted igniter was used for the ignition. For multi-droplet combustion experiments, either the droplet at the center or that on the right-hand-side was ignited.  \par

For single droplet of biodiesel doped with graphene oxide, after about 20\% of the droplet lifetime, heterogeneous nucleation occurred at the liquid and graphene oxide or liquid and fiber interface, leading to atomizations. This caused coupled and lateral movements of the droplet and its flame. Towards the end of the droplet lifetime, these atomizations intensified, significantly altering the droplet and flame movements. Though adding graphene oxide did not substantially influence the above dynamics, blending ethanol with biodiesel drastically influenced the dynamics. Specifically, immediately after the droplet was lit, nucleation at liquid-liquid interface (namely homogeneous nucleation) occurred leading to intense atomizations. The rms of the spatially integrated flame chemiluminescence and the lateral position of the chemiluminescence centroid were calculated for all tested fuels and for scenarios with no atomization, heterogeneous atomizations, and homogeneous atomizations. It was concluded that the rms of flame oscillations increases in the presence of atomizations, with that during homogeneous atomizations being at least three times larger than that during heterogeneous atomizations. The spectral analysis of biodiesel droplet combustion dynamics suggested that the heterogeneous atomization led to large amplitude oscillations of the flame chemiluminescence and its centroid lateral movements at about 10~Hz. Similar to biodiesel droplets, ethanol blended with biodiesel droplets also featured large amplitude oscillations at about 10~Hz and towards the end of the droplet lifetime. However, ethanol blended with biodiesel droplets featured large amplitude flame chemiluminescence and its centroid lateral oscillations at frequencies less than 50~Hz and immediately after the droplet was lit. Such dynamics of ethanol blended with biodiesel droplet combustion was not significantly influenced by the addition of graphene oxide. The low frequency dynamics observed for single droplet combustion was also present for multi-droplet combustion. The most dominant frequency of flame chemiluminescence oscillations was about 10--15 Hz for all test conditions and for both single and multi-droplet configurations. \par

The impact of atomization on the flame spread in a multi-droplet configuration was studied. Four modes of flame spread were identified. The first two modes were identical to those reported in past studies and were driven by the diffusion of heat and mass. However, for the third and fourth modes, the flame spread was driven by the heterogeneous and homogeneous atomizations, respectively. The results showed that, for biodiesel doped with graphene oxide as well as for relatively large inter-droplet spacing, the flame spreads from the main to the side droplet due to heterogeneous atomizations. For ethanol blended with doped biodiesel, independent of the inter-droplet spacing, the flame spread occurred due to homogeneous atomizations. For all of the above identified modes, the flame spread rate (the inter-droplet spacing multiplied by the droplet initial diameter divided by the flame spread time) was estimated. The flame spread time was defined as the difference between the time that the main droplet (i.e. the ignited droplet) was enveloped by the flame and the time that a side droplet was enveloped by the flame. It was observed that, while the first mode was not substantially influenced by the inter-droplets spacing, increasing this parameter slightly increased the flame spread rate from about 6 to 8~$\mathrm{mm^2/s}$ for the second mode. This was argued to be linked to elongated heat up period of the side droplet and (potentially) an increased amount of fuel vapor around the side droplet. For the mode associated with the heterogeneous atomizations, the flame spread rate was smaller than other modes and for similar inter-droplet spacing, which was linked to the random ejection of baby droplets and loss of combustion heat during the atomizations. Compared to this, for the mode associated with homogeneous atomizations, the flame spread rate was several times larger than the others (10--40~$\mathrm{mm^2/s}$). It was shown that, compared to the heterogeneous atomizations, the flame spread rate featured an increasing trend with the rms of the flame chemiluminescence centroid lateral movement for homogeneous atomizations. Comparing the results of the present study with those of past investigations, it was concluded that the impacts of inter-droplet spacing and GO addition on the flame spread rate are relatively small for low volatility fuels (such as biodiesel). However, adding volatile components to biodiesel significantly increases the flame spread rate.

\section*{Acknowledgments}
The authors are grateful for the financial support from the Natural Sciences Engineering Research Council of Canada and Zentek through the Alliance grant ALLRP 567111-21.

\bibliography{bibliography}

\end{document}